\pdfoutput=1
\documentclass[hidetodos]{jkpaper}
\usepackage{tikz}
\usepackage{aas_macros}
\usetikzlibrary{intersections}
\usetikzlibrary{patterns}
\usetikzlibrary{decorations.pathreplacing}
\newcommand{\pgfextractangle}[3]{%
    \pgfmathanglebetweenpoints{\pgfpointanchor{#2}{center}}
                              {\pgfpointanchor{#3}{center}}
    \global\let#1\pgfmathresult  
}

\title{The Holographic Dual of the\texorpdfstring{\\}{  }Entanglement Wedge Symplectic Form}
\author{Josh Kirklin}
\email{jjvk2@cam.ac.uk}
\institution{Department of Applied Mathematics and Theoretical Physics, Centre for Mathematical Sciences, University of Cambridge, Cambridge, UK}

\abstr{%
    In this paper, we find the boundary dual of the symplectic form for the bulk fields in any entanglement wedge. The key ingredient is Uhlmann holonomy, which is a notion of parallel transport of purifications of density matrices based on a maximisation of transition probabilities. Using a replica trick, we compute this holonomy for curves of reduced states in boundary subregions of holographic QFTs at large $N$, subject to changes of operator insertions on the boundary. It is shown that the Berry phase along Uhlmann parallel paths may be written as the integral of an abelian connection whose curvature is the symplectic form of the entanglement wedge. This generalises previous work on holographic Berry curvature.
}

\bibliography{refs}

\begin{document}

\maketitleandtoc

\section{Introduction}
\label{Section: Introduction}

\begin{figure}
    \centering
    \begin{tikzpicture}[thick,scale=1.3]
        \path (1.5,0) arc (0:85:1.5 and 0.5) coordinate (B);
        \path (1.5,0) arc (0:-85:1.5 and 0.5) coordinate (F);

        \fill[red!20] (F) .. controls (0.4,-0.3) and (0.4,0.3) .. (B) -- (1.5,1.3) -- (1.5,-1.3) -- (F);
        \fill[blue!20] (F) .. controls (0.7,0) and (1,0.2) .. (1.5,1.3) -- (1.5,-1.3) -- (F);
        \path[pattern=north west lines, pattern color=red!70] (F) .. controls (0.4,-0.3) and (0.4,0.3) .. (B) arc (85:-85:1.5 and 0.5);

        \draw[gray,dotted] (-0.9,-1.5) arc (180:0:1.2 and 0.5);
        \draw[gray] (-0.9,1.7) -- (-0.9,-1.5) arc (-180:0:1.2 and 0.5) -- (1.5,1.7);
        \draw[gray] (0.3,1.7) ellipse (1.2 and 0.5);

        \draw (1.5,-1.3) -- (1.5,1.3);

        \draw (F) -- (1.5,-1.3);
        \draw[dotted] (B) -- (1.5,-1.3);
        \draw (F) .. controls (0.7,0) and (1,0.2) .. (1.5,1.3);
        \draw[dotted] (B) -- (1.5,1.3);

        \draw[red,line width=1.5pt] (F) .. controls (0.4,-0.3) and (0.4,0.3) .. (B);
        \node at (0.1,0) {$\Upsilon$};

        \draw[dashed,blue,line width=1pt] (1.5,0) arc (0:85:1.5 and 0.5);
        \draw[blue, line width=1.5pt] (1.5,0) arc (0:-85:1.5 and 0.5);

        \node[right] at (1.5,0) {$A$};

        \draw[line width=0.7pt,<-] (0.6,0.2) .. controls (0.5,0.4) and (0,0.4) .. (-0.1,0.4) node[left] {$\Sigma$};

        \draw [decorate,decoration={brace,amplitude=10pt},xshift=-4pt,yshift=0pt] (2.2,1.3) -- (2.2,-1.3) node [right,midway,align=left,xshift=1em] {entanglement\\wedge of $A$};
    \end{tikzpicture}
    \caption{The entanglement wedge of a boundary subregion $A$ is defined as the domain of dependence of a partial Cauchy surface $\Sigma$ interpolating between $A$ and its associated HRT surface $\Upsilon$. (The colour scheme here will be used throughout this paper. Blue colouring indicates something on the boundary, whereas red colouring indicates something in the bulk.)}
    \label{Figure: entanglement wedge}
\end{figure}
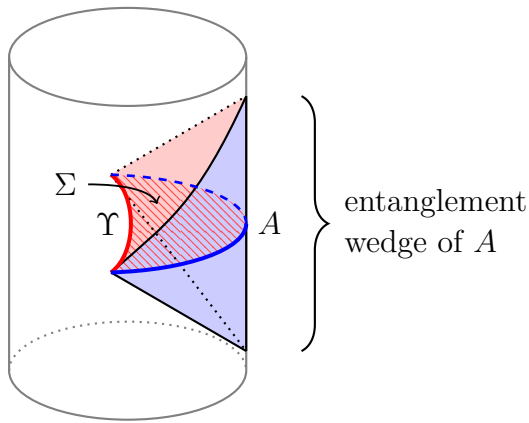

Most research stemming from the discovery of the AdS/CFT correspondence~\cite{Maldacena:1997re,Witten:1998qj} can be loosely sorted into two categories. The first involves using the duality to translate a hard question about quantum field theory into an easier one about gravity, or vice-versa. This translation makes use of the so-called holographic dictionary, i.e.\ the collection of 1-to-1 maps between concepts in the bulk gravity theory and the boundary field theory. But many pages of the dictionary remain empty, and the second category of research endeavours to fill these pages with new entries, in order to both deepen our understanding of holography, and widen the scope for its potential applications. In recent years a coherent picture of a particular section of the dictionary, under the heading `subregion duality', has emerged~\cite{Ryu:2006bv,Hubeny:2007xt,VanRaamsdonk:2009ar,Bousso:2012sj,Czech:2012bh,Bousso:2012mh,Headrick:2014cta,Almheiri:2014lwa,Jafferis:2015del,Dong:2016eik,Donnelly:2016qqt,Faulkner:2017vdd}. The entries in this section make precise the relationship between boundary locality and bulk locality by identifying properties of a given subregion of the boundary with those of an associated subregion of the bulk. The current consensus is that the bulk dual of a boundary subregion with Cauchy surface $A$ is its `entanglement wedge', which is the domain of dependence of a codimension 1 surface in the bulk joining $A$ with its Hubeny-Rangamani-Ryu-Takayanagi (HRT) surface (i.e.\ the codimension 2 surface homologous to $A$ in the bulk with extremal area). The standard depiction of the entanglement wedge is given in Figure~\ref{Figure: entanglement wedge}.

This paper makes an argument for a new entry in this section of the dictionary. To explain our new entry, consider the classical large $N$ limit of the bulk gravity theory. Such a limit should permit a classical Hamiltonian description, including a phase space whose points correspond to the different possible classical field configurations. The phase space comes equipped with a symplectic structure, i.e.\ a closed non-degenerate 2-form on that space, which, physically speaking, enables one to list all the pairs of conjugate variables in the theory, and to compute Poisson brackets of functions of these variables. In principle, this information is all one needs to construct a (low energy) perturbative Hilbert space. 

A popular and versatile construction of the classical phase space of a theory of fields, known as the covariant phase space formalism~\cite{Peierls:10.2307/99080,DeWitt:1985bc,DeWitt:2003pm,Bergmann:PhysRev.89.4,Crnkovic,300yearscrnkovicwitten,Zuckerman:1989cx,ashtekar1982,Lee:1990nz,Brown:1992br,Marolf:1993zk,Iyer:1994ys,Wald:1999wa,Barnich:2001jy,Hollands:2006zu}, is as follows. One considers the space of all possible on-shell field configurations $\phi$. Vector fields $\delta\phi$ on this space may be viewed as linearised on-shell field variations. The change in the Lagrangian density $L$ (which is a field-dependent spacetime top form) corresponding to such a variation may always be written to linear order in $\delta\phi$ as
\begin{equation}
    \delta L = L[\phi+\delta\phi] - L[\phi] = \delta\phi \cdot E + \dd{\theta}.
    \label{Equation: Lagrangian variation}
\end{equation}
The $\cdot$ denotes a sum over fields. $E=E[\phi]=0$ are the equations of motion, and are obeyed for the configurations we are considering by construction, so the variation of the Lagrangian is just equal to the exterior derivative of the form $\theta=\theta[\phi,\delta\phi]$. Consider now two different field variations $\delta_1\phi,\delta_2\phi$. Since these are vector fields on phase space, one may compute their Lie bracket $[\delta_1\phi,\delta_2\phi]$, which gives a third field variation. Let us define
\begin{equation}
    \omega[\phi,\delta_1\phi,\delta_2\phi] = \delta_1(\theta[\phi,\delta_2\phi])-\delta_2(\theta[\phi,\delta_1\phi])-\theta[\phi,[\delta_1\phi,\delta_2\phi]].
    \label{Equation: omega}
\end{equation}
If $\Sigma$ is a Cauchy surface, then 
\begin{equation}
    \Omega = \int_\Sigma\omega
\end{equation}
is an antisymmetric bilinear functional of the field variations $\delta_1\phi,\delta_2\phi$, and hence is a 2-form on phase space. One may further show that it is closed. In the covariant phase space formalism, $\Omega$ is the symplectic structure. If one wants a symplectic structure for the degrees of freedom in a subregion, one may generalise the above so that $\Sigma$ is only a partial Cauchy surface, and then $\Omega$ would be the symplectic structure for the domain of dependence of $\Sigma$.

Unfortunately this recipe has a well-known ambiguity in the presence of boundaries. The form $\theta$ is only defined up to the addition of an exact form, which means that $\Omega$ is only defined up to the addition of an integral over $\partial\Sigma$. Let us now make a distinction between what one might call `external' and `internal' boundaries. An external boundary is one beyond which there is no physics -- for example, asymptotic infinity is an external boundary. On the other hand, an internal boundary is the imaginary divider between whichever subregion we wish to consider and the rest of the system. If $\Sigma$ is a complete Cauchy surface, then $\partial\Sigma$ is entirely external. However, when $\Sigma$ is only a partial Cauchy surface, $\partial\Sigma$ may contain internal components. The ambiguity may be brought under control at external boundaries (e.g.~\cite{Wald:1999wa,Harlow:2019yfa}), and at internal boundaries in non-gravitational theories (e.g.~\cite{Kirklin:2019xug}), but a resolution at internal boundaries in a gravitational theory has been unclear. As a consequence, the ambiguity has inevitably been lurking in the background whenever these techniques have been used to understand locality in gravity. The ambiguous boundary terms are particularly important whenever one is trying to understand the role of edge modes and gauge transformations in the factorisation of the gravitational Hilbert space~\cite{Donnelly:2014fua,Donnelly:2016auv,Donnelly:2016rvo,Speranza:2017gxd,Camps:2018wjf,Dong:2018seb}, and in the closely related study of black hole soft hair~\cite{Donnay:2015abr,Hawking:2016msc,Hawking:2016sgy,Donnay:2016ejv,Haco:2018ske,Haco:2019ggi}.

\begin{figure}
    \centering
    \begin{tikzpicture}[scale=0.9]
        \node at (-0.3,0) {\Large$\mathcal{H}$};
        \begin{scope}[shift={(2.8,-2.5)}]
            \draw[thick] (0,0) circle (2);
            \foreach \x in {1,...,12} {
                \draw[purple!50!black,line width=1.5pt,->] (\x*30+25:2) arc (\x*30+25:\x*30+55:2);
            }
            \foreach \x in {1,...,12} {
                \fill[purple] (\x*30+10:2) circle (0.1);
            }
            \foreach \x in {1,...,10} {
                \node at (\x*30+10:2.6) {$\ket{\psi_{\x}}$};
            }
            \node at (340:2.6) {$\cdots$};
            \node at (10:2.6) {$\ket{\psi_n}$};
        \end{scope}
        \begin{scope}[shift={(11,-2.5)}]
            \draw[thick] (0,0) circle (2);
            \foreach \x in {1,...,60} {
                \fill[purple] (\x*6+10:2) circle (0.1);
            }co
        \end{scope}
        \draw[line width=1.5pt,->] (6.4,-2.5) -- (8.2,-2.5) node[above,midway] {$\lim_{n\to\infty}$};
    \end{tikzpicture}
    \caption{To find the Berry phase of a curve of normalised states in a Hilbert space $\mathcal{H}$, one picks a sequence of states $\ket{\psi_i}$ along that curve, and computes the product of the successive transition amplitudes between these states. One then takes the limit in which the sequence of states densely covers the curve.}
    \label{Figure: Berry phase}
\end{figure}
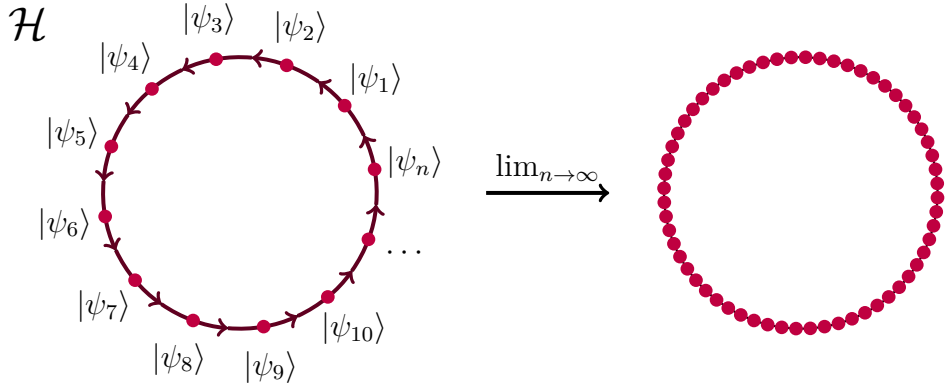

In the case where $\Sigma$ is a complete Cauchy surface for a bulk asymptotically AdS spacetime in a holographic theory, the boundary dual to $\Omega$ has recently been understood in terms of the Berry curvature~\cite{Pancharatnam1956,Berry1984,Simon1983,Wilczek1984,Aharonov:1987gg,Bhandari1988} of the boundary Hilbert space. To remind the reader of the definition of Berry curvature, consider a closed curve $C:S^1\to\mathcal{H}$ of normalised states in a Hilbert space $\mathcal{H}$, and suppose we choose a sequence of $n$ states ordered along this curve, $\ket*{\psi_1},\ket*{\psi_2},\dots,\ket*{\psi_n}$. Consider a limit in which $n\to\infty$ and the states $\ket{\psi_i}$ densely cover the curve $C$, as shown in Figure~\ref{Figure: Berry phase}. Then one may show that
\begin{equation}
    \braket*{\psi_1}{\psi_n}\braket*{\psi_n}{\psi_{n-1}}\dots\braket*{\psi_3}{\psi_2}\braket*{\psi_2}{\psi_1} \longrightarrow \exp(i\gamma),
\end{equation}
where $\gamma = \oint_C a$, and 
\begin{equation}
    a = i\mel*{\psi}{\mathrm{d}}{\psi}
\end{equation}
is a real 1-form on Hilbert space. In other words, upon traversing the curve $C$, the state of the system picks up a phase shift given by $\gamma$. This is the Berry phase.\footnote{The original definition of Berry phase in terms of the eigenstates of a slowly varying Hamiltonian is a special case of the one given here. This simpler and more general definition is sufficient for our purposes.} The map $\ket{\psi}\to e^{if}\ket{\psi}$, where $f$ is a real function on Hilbert space, is a gauge transformation that leaves the Berry phase unchanged. However, under this transformation we do have $a\to a-\dd{f}$. In other words $a$ transforms like a $U(1)$ connection; it is called the Berry connection. The curvature of the Berry connection (called the Berry curvature) is gauge-invariant, and is given by the formula
\begin{equation}
    \dd{a} = i\dd{\bra*{\psi}}\wedge\dd{\ket*{\psi}}.
\end{equation}

Returning to the holographic context, one may construct boundary states $\ket{\lambda}$ by inserting operators in a Euclidean path integral. The parameters $\lambda$ are the coefficients of these operator insertions, and set the boundary conditions for the bulk fields; in the classical limit there is a 1-to-1 map between the boundary conditions $\lambda$ and bulk field configurations $\phi$. It was shown in~\cite{Belin:2018fxe,Belin:2018bpg} that the bulk symplectic form is equal to the pullback of the boundary Berry curvature through this map. In light of subregion duality, an immediate question presents itself: is there a generalisation of this result to subregions? The purpose of this paper is to answer this question in the affirmative.

On the boundary side, we consider states $\ket{\lambda}$ reduced to a fixed subregion $A$. To be precise, this means the reduced density matrix
\begin{equation}
    \rho(\lambda) = \tr_{\bar{A}}\ket{\lambda}\bra{\lambda},
\end{equation}
where $\tr_{\bar{A}}$ denotes a trace over the part of the boundary Hilbert space containing the degrees of freedom in $\bar{A}$, the complement of $A$. Because of entanglement between $A$ and $\bar{A}$, $\rho(\lambda)$ is in general a mixed state, but Berry phases are only defined for pure states. Thus, we will need a generalisation of Berry phase. The generalisation we use is due to Uhlmann~\cite{Uhlmann:sp,Uhlmann1991,Uhlmann1992,Braunstein}, and is based on a maximisation of transition probabilities between purifications of density matrices. This leads to a notion of holonomy in the space of purifications, and will be described in more detail in Section~\ref{Section: Uhlmann holonomy}. For now, suffice it to say that to any closed curve of density matrices of the above form one may associate a phase shift, which we will refer to as the Uhlmann phase, and which may be written in terms of the integral of a connection around that curve. The curvature of this connection is the boundary quantity that we are interested in.

On the bulk side, one may compute the symplectic form $\Omega$ of the entanglement wedge of $A$, using the covariant phase space recipe. Such a symplectic form is subject to the boundary ambiguity mentioned above at the HRT surface, which is an internal boundary. There is a particular way to resolve this ambiguity which will be described in this paper.

Our claim is that this now unambiguous entanglement wedge symplectic form is exactly dual to the curvature of the Uhlmann phase of $A$. This generalises in a very natural way the result of~\cite{Belin:2018fxe,Belin:2018bpg}. At the same time, it fulfils a more general principle for resolving the boundary ambiguity in the symplectic form. In a certain sense, the boundary ambiguity is representative of the following fact. When one divides space into two subregions joined by a common boundary, one must make a choice about the degrees of freedom that lie on that boundary. In particular, one must decide which of the two subregions each such degree of freedom should be associated with, and in principle, without additional constraints, one is free to make this decision however one likes. However, the holographic context is an additional constraint. There is only one way to sort the degrees of freedom on the boundary in a way that is consistent with subregion duality. The resolution of the ambiguity presented in this paper is thus exactly the one implied by the holographic correspondence.

It is worth pointing out that Uhlmann holonomy is a direct probe of entanglement. Thus, our result adds to the long-growing list of evidence that entanglement is a key ingredient in the emergence of bulk physics. Related ideas concerning entanglement and holonomy have appeared in~\cite{Czech:2017zfq,Czech:2018kvg,Czech:2019vih}. However, in those papers the authors were chiefly concerned with deformations of boundary subregions in the presence of fixed sources, whereas here we fix the boundary subregion and vary the sources. A unified picture of the results in those papers and the present one is likely to exist, and a potential approach to this will be presented in Section~\ref{Section: edge modes}.

This paper begins with a brief review in Section~\ref{Section: Uhlmann holonomy} of Uhlmann holonomy, and the related notions of fidelity and parallel purifications. In Section~\ref{Section: Holographic}, we describe the construction of holographic states reduced to a subregion, and use a replica trick to find a convenient formula for the fidelity of such states. This allows us to find a sequence of parallel purifications along any given curve of such states, and to compute the Uhlmann phase of the curve. In Section~\ref{Section: Symplectic form of the entanglement wedge}, we demonstrate the equivalence between the curvature of this phase and the symplectic form of the entanglement wedge, and describe the way in which the boundary ambiguity is resolved. We also comment upon the existence of edge modes corresponding to deformations of the HRT surface. We conclude the paper in Section~\ref{Section: Conclusion} with some brief discussion and comments on possible applications.

\section{Uhlmann holonomy and fidelity}
\label{Section: Uhlmann holonomy}

In this section we will briefly describe and motivate Uhlmann holonomy. The proofs of several claims below can be found in the literature, e.g.\ in~\cite{Uhlmann:sp,Uhlmann1991,Uhlmann1992,Braunstein}.

Suppose $\rho$ is a density matrix acting on a Hilbert space $\mathcal{H}$. A purification of $\rho$ is a pure state $\ket{\psi}$ (which we will assume for simplicity is normalised) in an extended Hilbert space $\mathcal{H}\otimes\mathcal{H}'$ such that
\begin{equation}
    \rho = \tr'\ket{\psi}\bra{\psi},
    \label{Equation: purification}
\end{equation}
where $\tr'$ denotes a partial trace over $\mathcal{H}'$. The auxiliary space $\mathcal{H}'$ can be any Hilbert space one wants, and for each choice there can exist many possible purifications of a given density matrix. 

Let us suppose that by measuring a system at two different times we determine that it is initially in one state $\rho_1$, and then subsequently in a different state $\rho_2$. Let us also assume also that these density matrices arise as reductions of some pure states $\ket*{\psi_1},\ket*{\psi_2}$ in an extended system, but that we know nothing else about those states. Despite our ignorance about each of the purifications by themselves, we can say something about the relationship between them. In particular, the transition probability for $\ket*{\psi_1}\to\ket*{\psi_2}$ is
\begin{equation}
    |\braket*{\psi_2}{\psi_1}|^2.
    \label{Equation: transition probability}
\end{equation}
The key idea of Uhlmann is to assume that $\ket*{\psi_1},\ket*{\psi_2}$ maximise this probability. If we are in a classical regime in which the transition probability distribution is sharply peaked, then on statistical grounds this assumption is a good approximation. We call purifications which satisfy this maximisation condition `parallel'.

The following relation holds if and only if $\ket*{\psi_1},\ket*{\psi_2}$ are parallel purifications:
\begin{equation}
    |\braket*{\psi_2}{\psi_1}| = \tr(\sqrt{\sqrt{\rho_1}\rho_2\sqrt{\rho_1}}).
    \label{Equation: Uhlmann's theorem}
\end{equation}
Here the square root of a positive Hermitian operator is just defined in terms of its spectrum. The quantity on the right-hand side is known as the fidelity of $\rho_1,\rho_2$, and it is the square root of a generalisation of transition probability to mixed states. This result, sometimes known as Uhlmann's theorem, provides a useful criterion for determining when two purifications are parallel, and we will make use of it in this paper. It is proven, for example, in~\cite{Jozsa:1994}.

Parallel purifications are not unique, because if $\ket*{\psi_1},\ket*{\psi_2}$ are purifications satisfying \eqref{Equation: Uhlmann's theorem}, then so are
\begin{equation}
    e^{if_1}(I\otimes U)\ket*{\psi_1},\quad e^{if_2}(I\otimes U)\ket*{\psi_2},
    \label{Equation: parallel purification change}
\end{equation}
where $f_1,f_2$ are any two real numbers, and $U$ is any unitary operator acting on the auxiliary Hilbert space $\mathcal{H}'$. Indeed, the transition probability \eqref{Equation: transition probability} is unaffected if we change the states in this way. By choosing $f_1,f_2,U$ appropriately, one can in fact obtain all possible parallel purifications of $\rho_1,\rho_2$.

Suppose now that we have a closed curve $C$ of density matrices acting on $\mathcal{H}$, and let $\rho_1,\rho_2,\dots,\rho_n$ be a sequence of $n$ density matrices ordered along this curve. Let us assume that we have a sequence of states $\ket*{\psi_1},\ket*{\psi_2},\dots,\ket*{\psi_n}$ in an extended Hilbert space $\mathcal{H}\otimes\mathcal{H}'$ such that each $\ket*{\psi_i}$ purifies $\rho_i$, and such that each consecutive pair $\ket*{\psi_i},\ket*{\psi_{i+1}}$ of states is parallel. Consider the limit in which $n\to \infty$ and the density matrices $\rho_i$ densely cover the curve $C$. We will assume that one can choose the phases of the purifications $\ket*{\psi_i}$ is such a way that they converge in this limit to a dense cover of a curve $\tilde{C}$ in $\mathcal{H}\otimes\mathcal{H}'$. Then we say that $\tilde{C}$ is a parallel lift of $C$. 

One can directly construct parallel lifts for curves of faithful states\footnote{Faithful states are those with an invertible density matrix. All the states we consider in this paper are either pure or faithful.} in the following way. Let $t\in[0,1]$ be a parameter along $C$ such that $\rho=\rho(t)$ is the density matrix at $t$, and consider the differential equation
\begin{equation}
    \dv{t}\ket*{\psi(t)} = \qty(\int_0^\infty \dd{s} e^{-s\rho(t)}\dot\rho(t)e^{-s\rho(t)} \otimes I)\ket*{\psi(t)}.
\end{equation}
The integral on the right-hand side is convergent because $\rho$ is a positive operator. Along with an initial condition $\ket*{\psi(0)}$ that purifies $\rho(0)$, this equation may be solved to give a curve in $\mathcal{H}\otimes\mathcal{H}'$, and it may be verified that this curve is a parallel lift of $C$. Of course, this is not the unique parallel lift of $C$, as \eqref{Equation: parallel purification change} is still allowed, the continuous version of which is
\begin{equation}
    \ket*{\psi(t)} \to e^{if(t)} (I\otimes U)\ket*{\psi(t)},
    \label{Equation: different Uhlmann curves}
\end{equation}
for some real function $f:[0,1]\to \RR$, and constant unitary $U$ acting on $\mathcal{H}'$. If we fix the initial condition $\ket*{\psi(0)}$, then one may set $U=1,f(0)=0$, and the space of all allowed parallel lifts of $\rho(t)$ is given by curves of the form $e^{if(t)} \ket*{\psi(t)}$. In this way, parallel lifts of density matrices provide a notion of parallel transport of purifications modulo phase shifts. This is Uhlmann holonomy.

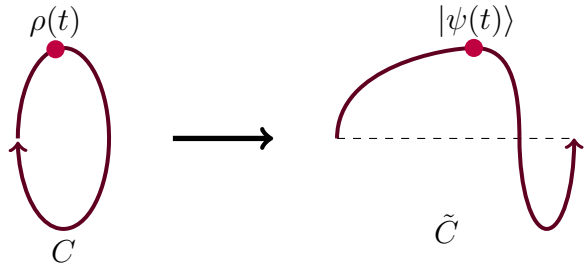
\begin{figure}
    \centering
    \begin{tikzpicture}[scale=1.2]
        \draw[purple!50!black, line width=1.5pt,->] (0.5,0) arc (180:100:0.5 and 1) coordinate (A) arc (100:-178:0.5 and 1);
        \fill[purple] (A) circle (0.1);
        \node[above] at (A) {$\rho(t)$};
        \node[below] at (1,-1) {$C$};

        \draw[line width=2pt,->] (2.2,0) -- (3.3,0);

        \begin{scope}[shift={(4,0)}]
            \draw[dashed] (0,0) -- (2.6,0);
            \draw[purple!50!black, line width=1.5pt,->] (0,0) .. controls (0,0.8) and (1.2,1) .. (1.5,1) .. controls (1.8,1) and (2,0.8) .. (2,0) arc (-180:0:0.3 and 1);
            \fill[purple] (1.5,1) circle (0.1);
            \node[above] at (1.5,1) {$\ket*{\psi(t)}$};
            \node at (1.2,-1) {$\tilde C$};
        \end{scope}
    \end{tikzpicture}
    \caption{A curve $C$ of density matrices gives rise to a parallel lift $\tilde{C}$ of purifications. Even if $C$ is closed, it may be impossible for $\tilde{C}$ to be, because of curvature in the Uhlmann holonomy.}
    \label{Figure: parallel lift}
\end{figure}

It is worth noting that although $C$ may be a closed curve, in general its parallel lift $\tilde{C}$ is not (even up to phase shifts), as shown in Figure~\ref{Figure: parallel lift}. This is because a purification will sometimes not return to itself upon being parallelly transported around the curve. Indeed, we are only guaranteed 
\begin{equation}
    \ket*{\psi(1)}=(I\otimes X) \ket*{\psi(0)},
    \label{Equation: Uhlmann holonomy}
\end{equation}
where $X$ is a unitary operator acting on $\mathcal{H}'$. In other words, there is non-trivial curvature in the Uhlmann holonomy. This is due to entanglement between $\mathcal{H}$ and $\mathcal{H}'$. We cannot eliminate $X$ by doing a transformation of the form \eqref{Equation: different Uhlmann curves}, because the operator $U$ must be constant, and so acts in the same way on both sides of \eqref{Equation: Uhlmann holonomy}.

Consider now the quantity $\gamma$ defined by
\begin{equation}
    e^{i\gamma} = \lim_{n\to\infty} \braket*{\psi_1}{\psi_n}\braket*{\psi_n}{\psi_{n-1}}\dots\braket*{\psi_3}{\psi_2}\braket*{\psi_2}{\psi_1},
\end{equation}
i.e.\ the Berry phase along $\tilde{C}$. This is clearly invariant under a change in parallel purifications $\ket*{\psi_i}\to e^{if_i} (I\otimes U)\ket*{\psi_i}$, and so is uniquely defined for any closed curve $C$ of density matrices. We will refer to it as the Uhlmann phase of such a curve. For the special case of a curve of density matrices representing pure states, the Uhlmann phase reduces to the Berry phase.

Uhlmann holonomy may be viewed as a map from a curve $C$ in the space of density matrices, and an initial purification $\ket*{\psi}\in\mathcal{H}\otimes\mathcal{H}'$, to the unitary operator $X$ in \eqref{Equation: Uhlmann holonomy} (modulo phase shifts). Since the group of unitary operators acting on $\mathcal{H}'$ is in general non-abelian, it is clear that the Uhlmann holonomy is also non-abelian. However, in the classical regime the effects of operator ordering become subleading, and so we can expect to be able to approximate $e^{i\gamma}$ as the holonomy of an abelian $U(1)$ connection on the space of density matrices.\footnote{The reader may be concerned with the imprecision in the justification of this statement. We will only comment that, in the holographic case described in this paper, $e^{i\gamma}$ does indeed take this form in the large $N$ limit.} It is the curvature of this connection that will interest us the most in the next section.

\section{Holographic Uhlmann holonomy}
\label{Section: Holographic}
In this Section, we will calculate the Uhlmann phase in a holographic theory. To start, we will define the states of interest, and find an expression for their fidelity. This will then allow us to invoke Uhlmann's theorem to find parallel purifications, and from there compute the Uhlmann holonomy.

Consider a $d$-dimensional holographic CFT. Let us define the following class of states:
\begin{equation}
    \ket*{\lambda} = \mathrm{T} \exp(-\int_{\tau<0} \dd{\tau}\dd[d-1]{x} \lambda(\tau,x)\cdot\mathcal{O}(\tau,x)) \ket{0}.
    \label{Equation: holographic state with sources}
\end{equation}
Here $\mathcal{O}$ is supposed to denote all possible single trace operators dual to bulk fields, and the parameter $\lambda$ is a function which sources these fields. The $\mathrm{T}$ denotes a Euclidean time $\tau$ ordering, and the remaining coordinates $x$ are the spatial ones. The state $\ket{0}$ on the right-hand side is usually the vacuum, whose wavefunctional is obtained by doing a Euclidean path integral over half of a $d$-sphere. It could also be a more complicated background state such as the thermofield double in two copies of the CFT, whose wavefunctional arises from a Euclidean path integral over $S^{d-1}\times I_{\beta/2}$, where $I_{\beta/2}$ is an interval of length $\beta/2$ and $\beta$ is the inverse temperature. Let us label the manifold on which this path integral is performed $\mathcal{M}^-$. The effect of the operator in front of $\ket{0}$ is to introduce additional sources on $\mathcal{M}^-$ in this path integral.

The dual states to $\ket*{\lambda}$ may be written
\begin{equation}
    \bra*{\lambda} = \bra{0}\mathrm{T} \exp(-\int_{\tau>0} \dd{\tau}\dd[d-1]{x} \lambda^*(-\tau,x)\cdot\mathcal{O}^\dagger(\tau,x)).
\end{equation}
One sees that $\bra*{\lambda}$ is related to $\ket*{\lambda}$ by a complex conjugation and reflection of the sources across $\tau=0$. Following~\cite{Belin:2018fxe}, we will refer to this transformation as $\mathbb{Z}_2+\mathcal{C}$, where $\mathbb{Z}_2$ refers to the time relection, and $\mathcal{C}$ refers to the complex conjugation. Let us call the reflected manifold on which this state is prepared $\mathcal{M}^+$.

The inner product of two such states may be evaluated as a path integral over the manifold obtained by gluing $\mathcal{M}^-$ and $\mathcal{M}^+$ at their boundaries. By `gluing', we mean identifying all the fields there, and summing over them. At leading order in the classical large $N$ limit, the holographic dictionary allows us to write this as
\begin{equation}
    \braket*{\lambda_2}{\lambda_1} = e^{-S(\lambda_1,\lambda_2)},
    \label{Equation: large N inner product}
\end{equation}
where $S(\lambda_1,\lambda_2)$ is the on-shell gravitational Euclidean action evaluated on a $(d+1)$-dimensional bulk with boundary conditions matching the sources $\lambda_1(\tau,x)$ for $\tau<0$ (i.e.\ on $\mathcal{M}^-$), and $\lambda_2^{\mathrm{T}*}(\tau,x)$ for $\tau>0$ (i.e.\ on $\mathcal{M}^+$). Here the superscript ${}^{\mathrm{T}}$ denotes a time reflection, so $\lambda_2^{\mathrm{T}*}(\tau,x) = \lambda_2^*(-\tau,x)$. In this paper, unless stated otherwise, all bulk actions are Euclidean. Figure~\ref{Figure: sourced states} contains an illustration of the path integrals for $\ket{\lambda}$, $\bra{\lambda}$ and $\braket{\lambda_2}{\lambda_1}$.

\begin{figure}
    \centering
    \begin{subfigure}{0.35\textwidth}
        \centering
        \begin{tikzpicture}[font=\large,scale=0.8]
            \fill[blue!15] (-2,0) arc (-180:0:2) arc (0:180:2 and 0.4);
            \fill[blue!20] (-2,0) arc (-180:0:2) arc (0:-180:2 and 0.4);
            \draw[thick] (-2,0) arc (-180:0:2 and 0.4);
            \draw[thick] (-2,0) arc (-180:0:2) arc (0:180:2 and 0.4);
            \fill (0.6,-1.1) circle (0.05) node[right] {$\lambda$};
            \node[below] at (0,-2.2) {$\ket{\lambda}$};
            \node[right] at (2,-1) {$\mathcal{M}^-$};
            \begin{scope}[shift={(0,3)}]
                \fill[blue!20] (-2,0) arc (180:0:2) arc (0:-180:2 and 0.4);
                \draw[black!65,dotted,thick] (-2,0) arc (180:0:2 and 0.4);
                \draw[thick] (-2,0) arc (180:0:2) arc (0:-180:2 and 0.4);
                \fill (0.6,0.8) circle (0.05) node[right] {$\lambda^{\mathrm{T}*}$};
                \node[below] at (0,-0.6) {$\bra{\lambda}$};
                \node[right] at (2,1) {$\mathcal{M}^+$};
            \end{scope}
            \draw[<->,thick] (-2.3,-0.3) .. controls (-3.5,0.2) and (-3.5,2.8) .. (-2.3,3.3) node[midway, right] {\normalsize$\,\mathbb{Z}_2+\mathcal{C}$};
        \end{tikzpicture}
        \caption{}
    \end{subfigure}
    \begin{subfigure}{0.35\textwidth}
        \centering
        \begin{tikzpicture}[font=\large]
            \fill[blue!20] (0,0) circle (2);
            \draw[blue!65,thick,dotted] (-2,0) arc (180:0:2 and 0.4);
            \draw[blue!65,thick] (2,0) arc (0:-180:2 and 0.4);
            \draw[thick] (0,0) circle (2);
            \fill (0.6,0.8) circle (0.05) node[right] {$\lambda_2^{\mathrm{T}*}$};
            \fill (-0.3,-1.1) circle (0.05) node[right] {$\lambda_1$};
            \node[below] at (0,-2.2) {$\braket{\lambda_2}{\lambda_1}$};
            \node[right] at (2,-1) {$\mathcal{M}^-$};
            \node[right] at (2,1) {$\mathcal{M}^+$};
        \end{tikzpicture}
        \caption{}
    \end{subfigure}
    \caption{\mbox{\textbf{(a)}\ The} states $\ket{\lambda}$ we are considering are prepared by a Euclidean path integral. The function $\lambda$ parametrises insertions of operators in this path integral. The dual states $\bra{\lambda}$ are prepared by doing the same path integral, but with everything acted upon by $\mathbb{Z}_2+\mathcal{C}$, where $\mathbb{Z}_2$ is Euclidean time reflection and $\mathcal{C}$ is complex conjugation. \mbox{\textbf{(b)}\ The} inner product of two such states is computed by doing a path integral on the manifold obtained by gluing together the two constituent manifolds along their boundaries. At large $N$, this manifold sets the boundary conditions for the bulk fields.}
    \label{Figure: sourced states}

    \vspace*{\floatsep}
    \vspace*{0.5in}

    \begin{tikzpicture}[font=\large]
        \fill[blue!15] (0,-0.1) ellipse (2 and 0.4);
        \draw[thick] (-2,-0.1) arc (180:0:2 and 0.4);
        \fill[blue!20] (-2,0.1) arc (180:0:2) -- (2,-0.1) arc (0:-180:2) arc (-180:-90:2 and 0.4) arc (-90:90:0.1) arc (-90:-180:2 and 0.4);
        \begin{scope}
            \clip (-2,0.1) arc (180:0:2) -- (2,-0.1) arc (0:-180:2) arc (-180:-90:2 and 0.4) arc (-90:90:0.1) arc (-90:-180:2 and 0.4);
            \draw[blue!65,thick] (2,0) arc (0:-90:2 and 0.4);
            \draw[blue!65,dotted,thick] (2,0) arc (0:88:2 and 0.4);
            \draw[black!65,dotted,thick] (-2,-0.1) arc (180:90:2 and 0.4) arc (-90:90:0.1) arc (90:180:2 and 0.4);
        \end{scope}
        \draw[thick] (-2,0.1) arc (180:0:2) -- (2,-0.1) arc (0:-180:2) arc (-180:-90:2 and 0.4) arc (-90:90:0.1) arc (-90:-180:2 and 0.4);
        \fill (-0.2,1.4) circle (0.05) node[right] {$\lambda^{\mathrm{T}*}$};
        \fill (-0.2,-1.7) circle (0.05) node[right] {$\lambda$};
        \node[below] at (1,-0.3) {$\bar{A}$};
        \node[below] at (-1,-0.45) {$A$};
        \node[below] at (0,-2.3) {$\rho(\lambda)$};
    \end{tikzpicture}
    \caption{The density matrix $\rho(\lambda)$ for a subregion $A$ may be prepared by taking the path integrals for $\ket{\lambda}$ and $\bra{\lambda}$, and gluing along $\bar{A}$, the complement of $A$.}
    \label{Figure: density matrix}
\end{figure}

We will use $\phi$ to denote the collection of bulk fields dual to boundary operators. It was shown in~\cite{Belin:2018fxe} that, at leading order in large $N$, the Berry curvature for normalised states of the above form matches exactly with the symplectic form of the bulk fields, where bulk field variations $\delta\phi$ are related with changes in boundary sources $\delta\lambda$ via the holographic dictionary.

Let us now fix a proper subregion $A\subset\partial\mathcal{M}^-$ of the boundary CFT at $\tau=0$, and factorise the boundary Hilbert space as $\mathcal{H}=\mathcal{H}_A\otimes\mathcal{H}_{\bar{A}}$, where $\mathcal{H}_{A,\bar{A}}$ are the Hilbert spaces for the degrees of freedom in $A,\bar{A}$ respectively, and $\bar{A}$ is the complement of $A$. The state in $A$ in the presence of the sources $\lambda$ is given by the density matrix
\begin{equation}
    \rho(\lambda) = e^{S(\lambda,\lambda)}\tr_{\bar{A}}\ket*{\lambda}\bra*{\lambda},
    \label{Equation: rho large N}
\end{equation}
i.e.\ by tracing over all degrees of freedom in $\bar{A}$. The prefactor involving $S(\lambda,\lambda)$ is necessary for the correct normalisation. This density matrix can be prepared by computing a path integral over the manifold obtained by gluing $\bar{A}\subset\partial\mathcal{M}^-$ to its mirror image under $\mathbb{Z}_2+\mathcal{C}$ in $\partial\mathcal{M}^+$. This is shown in Figure~\ref{Figure: density matrix}. The state $\rho(\lambda)$ is in general mixed due to the presence of entanglement between $\mathcal{H}_A$ and $\mathcal{H}_{\bar{A}}$ in $\ket{\lambda}$.

\subsection{Fidelity from a replica trick}
\label{Section: fidelity replica}

Suppose we have prepared two such states $\rho_1=\rho(\lambda_1), \rho_2=\rho(\lambda_2)$ in this way. In this section we will find an expression for the fidelity of these two states. The fidelity of holographic states has previously been considered in~\cite{Banerjee:2017qti,Alishahiha:2017cuk,Moosa:2018mik} and others.

Consider the operator $\sqrt{\sqrt{\rho_1}\rho_2\sqrt{\rho_1}}$, whose trace is the fidelity of $\rho_1$ and $\rho_2$. This operator is positive since 
\begin{equation}
    \sqrt{\rho_1}\rho_2\sqrt{\rho_1} = \qty(\sqrt{\rho_2}\sqrt{\rho_1})^\dagger\sqrt{\rho_2}\sqrt{\rho_1}.
\end{equation}
Furthermore, by \eqref{Equation: Uhlmann's theorem} the fidelity is equal to the inner product of two normalised states, and so the trace of this operator is less than or equal to 1. Note that in a QFT reduced states in proper subregions are faithful, so we can conclude that the operator is invertible and that all its eigenvalues lie strictly between 0 and 1.

Let us define the `replicated fidelity'\footnote{It is worth pointing out that this replica trick is different to the one employed in~\cite{Lashkari:2014yva}. In that paper, the `relative R\'enyi entropy' 
\begin{equation}
    S_k = \frac1{k-1}\log\tr(\qty(\rho_1^{\frac{1-k}{2k}}\rho_2\rho_2^{\frac{1-k}{2k}})^k)
\end{equation}
was the object considered. This is supposed to give the fidelity in the limit $k\to\frac12$. It would be interesting to see if one could use the relative R\'enyi entropy to get similar results to the ones found here.}
\begin{equation}
    F_k = \tr(\qty(\sqrt{\rho_1}\rho_2\sqrt{\rho_1})^k).
\end{equation}
By the above considerations, $F_k$ is analytic in $k$, and absolutely bounded by 1 for $\operatorname{Re} k\ge\frac12$. By Carlson's theorem, $F_k$ is therefore uniquely determined in this range by the values it takes on positive integers $k$. Our strategy to find the fidelity will be to compute $F_k$ for $k\in\ZZ_{>0}$, and then to analytically continue back to $k=\frac12$. This is easier than a direct calculation of the fidelity because the cyclic property of the trace means we can write
\begin{equation}
    F_k = \tr((\rho_1\rho_2)^k)\qquad \text{for } k\in\mathbb{Z}_{>0}.
    \label{Equation: replica fidelity}
\end{equation}

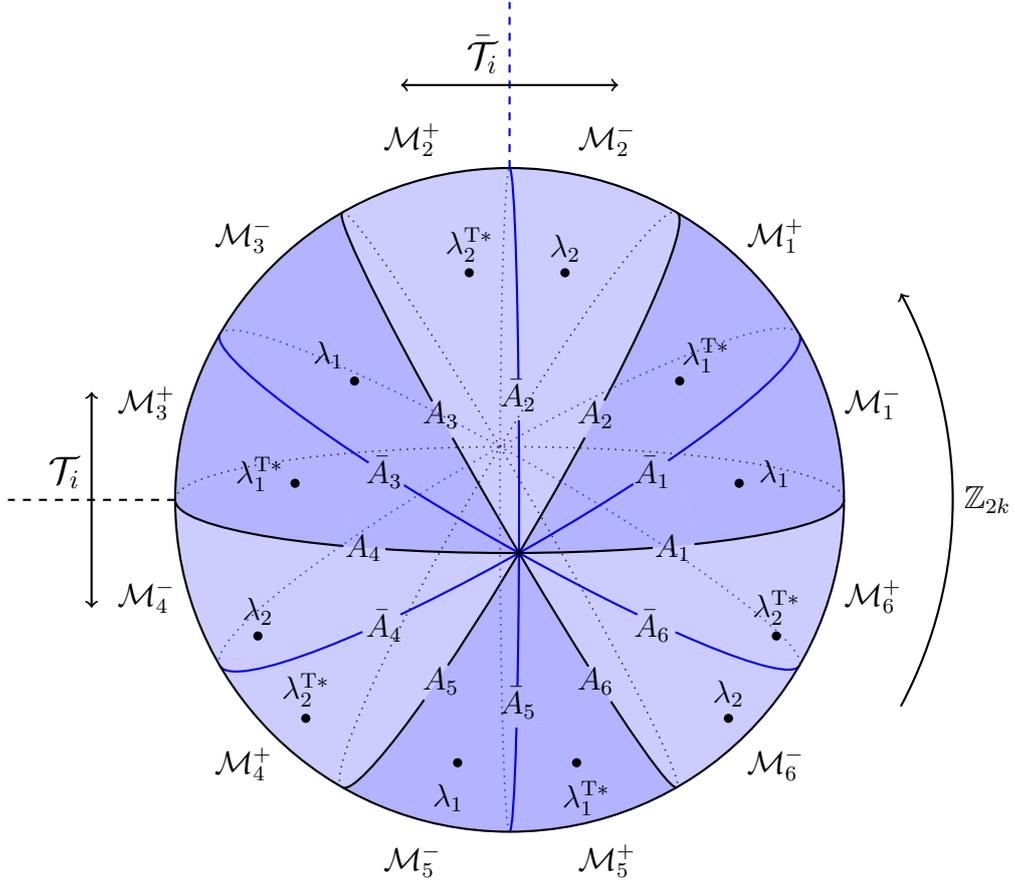
\begin{figure}
    \centering
    \begin{tikzpicture}[thick,scale=1.1]
        \pgfmathsetmacro{\bigradius}{4}
        \pgfmathsetmacro{\threeD}{0.65}
        \pgfmathsetmacro{\threeDtilt}{10}
        \fill[blue!20] (0,0) circle (\bigradius);
        \foreach \x in {3,7,11} {
            \pgfmathsetmacro{\theAngle}{\x*30+\threeDtilt}
            \pgfmathsetmacro{\littleradius}{\bigradius*\threeD*sin(\theAngle)/sqrt(\bigradius*\bigradius-\threeD*\threeD*cos(\theAngle)*cos(\theAngle))}
            \fill[blue!30,rotate=-\x*30] (0,-\bigradius) arc (-90:0:{\littleradius} and \bigradius) -- (-90+\theAngle:\threeD) -- (-120:\bigradius) arc (-120:-90:\bigradius);
        }
        \foreach \x in {1,5,9} {
            \pgfmathsetmacro{\theAngle}{\x*30+\threeDtilt}
            \pgfmathsetmacro{\littleradius}{\bigradius*\threeD*sin(\theAngle)/sqrt(\bigradius*\bigradius-\threeD*\threeD*cos(\theAngle)*cos(\theAngle))}
            \fill[blue!30,rotate=-\x*30] (0,-\bigradius) arc (-90:0:{\littleradius} and \bigradius) -- (-90+\theAngle:\threeD) -- (-60:\bigradius) arc (-60:-90:\bigradius);
        }
        \foreach \x in {0,2,4} {
            \pgfmathsetmacro{\theAngle}{\x*30+\threeDtilt}
            \pgfmathsetmacro{\littleradius}{\bigradius*\threeD*sin(\theAngle)/sqrt(\bigradius*\bigradius-\threeD*\threeD*cos(\theAngle)*cos(\theAngle))}
            \draw[blue,rotate=-\x*30] (0,-\bigradius) arc (-90:90:{\littleradius} and \bigradius);
        }
        \foreach \x in {1,3,5} {
            \pgfmathsetmacro{\theAngle}{\x*30+\threeDtilt}
            \pgfmathsetmacro{\littleradius}{\bigradius*\threeD*sin(\theAngle)/sqrt(\bigradius*\bigradius-\threeD*\threeD*cos(\theAngle)*cos(\theAngle))}
            \draw[rotate=-\x*30] (0,-\bigradius) arc (-90:90:{\littleradius} and \bigradius);
        }
        \begin{scope}
            \clip (-90+\threeDtilt:\threeD) circle (2.1);
            \fill[blue!20] (0,0) circle (\bigradius);
            \foreach \x in {3,7,11} {
                \pgfmathsetmacro{\theAngle}{\x*30+\threeDtilt}
                \pgfmathsetmacro{\littleradius}{\bigradius*\threeD*sin(\theAngle)/sqrt(\bigradius*\bigradius-\threeD*\threeD*cos(\theAngle)*cos(\theAngle))}
                \fill[blue!30,rotate=-\x*30] (0,-\bigradius) arc (-90:0:{\littleradius} and \bigradius) -- (-90+\theAngle:\threeD) -- (-120:\bigradius) arc (-120:-90:\bigradius);
            }
            \foreach \x in {1,5,9} {
                \pgfmathsetmacro{\theAngle}{\x*30+\threeDtilt}
                \pgfmathsetmacro{\littleradius}{\bigradius*\threeD*sin(\theAngle)/sqrt(\bigradius*\bigradius-\threeD*\threeD*cos(\theAngle)*cos(\theAngle))}
                \fill[blue!30,rotate=-\x*30] (0,-\bigradius) arc (-90:0:{\littleradius} and \bigradius) -- (-90+\theAngle:\threeD) -- (-60:\bigradius) arc (-60:-90:\bigradius);
            }
        \end{scope}
        \begin{scope}
            \clip (-90+\threeDtilt:\threeD) circle (1.6);
            \foreach \x in {0,2,4} {
                \pgfmathsetmacro{\theAngle}{\x*30+\threeDtilt}
                \pgfmathsetmacro{\littleradius}{\bigradius*\threeD*sin(\theAngle)/sqrt(\bigradius*\bigradius-\threeD*\threeD*cos(\theAngle)*cos(\theAngle))}
                \draw[blue,rotate=-\x*30] (0,-\bigradius) arc (-90:90:{\littleradius} and \bigradius);
            }
            \foreach \x in {1,3,5} {
                \pgfmathsetmacro{\theAngle}{\x*30+\threeDtilt}
                \pgfmathsetmacro{\littleradius}{\bigradius*\threeD*sin(\theAngle)/sqrt(\bigradius*\bigradius-\threeD*\threeD*cos(\theAngle)*cos(\theAngle))}
                \draw[rotate=-\x*30] (0,-\bigradius) arc (-90:90:{\littleradius} and \bigradius);
            }
        \end{scope}
        \foreach \x in {0,2,4} {
            \pgfmathsetmacro{\theAngle}{\x*30+\threeDtilt}
            \pgfmathsetmacro{\littleradius}{\bigradius*\threeD*sin(\theAngle)/sqrt(\bigradius*\bigradius-\threeD*\threeD*cos(\theAngle)*cos(\theAngle))}
            \draw[semithick,blue!80,dotted,rotate=-\x*30] (0,-\bigradius) arc (270:90:{\littleradius} and \bigradius);
        }
        \foreach \x in {1,3,5} {
            \pgfmathsetmacro{\theAngle}{\x*30+\threeDtilt}
            \pgfmathsetmacro{\littleradius}{\bigradius*\threeD*sin(\theAngle)/sqrt(\bigradius*\bigradius-\threeD*\threeD*cos(\theAngle)*cos(\theAngle))}
            \draw[semithick,black!80,dotted,rotate=-\x*30] (0,-\bigradius) arc (270:90:{\littleradius} and \bigradius);
        }
        \draw (0,0) circle (\bigradius);
        \foreach \x in {1,2,3,4,5,6} {
            \node at (-45+60*\x:4.5) {$\mathcal{M}^-_\x$};
            \node at (-15+60*\x:4.5) {$\mathcal{M}^+_\x$};
            \node[shift={(-90+\threeDtilt:\threeD)}] at (-30+60*\x:1.85) {$\bar{A}_\x$};
            \node[shift={(-90+\threeDtilt:\threeD)}] at (-60+60*\x:1.85) {$A_\x$};
        }
        \begin{scope}[shift={(-90+\threeDtilt:\threeD*0.8)},scale=1.1]
            \foreach \x in {2,4,6} {
                \fill (-40+60*\x:3) circle (0.05);
                \fill (-20+60*\x:3) circle (0.05);
                \node[above] at (-40+60*\x:3) {$\lambda_2$};
                \node[above] at (-20+60*\x:3) {$\lambda_2^{\mathrm{T}*}$};
                \fill (15+60*\x:2.5) circle (0.05);
                \fill (45+60*\x:2.5) circle (0.05);
                \node at (15+60*\x:2.9) {$\lambda_1$};
                \node at (45+60*\x:2.9) {$\lambda_1^{\mathrm{T}*}$};
            }
        \end{scope}
        \draw[thick,dashed,blue] (0,4) -- (0,6);
        \draw[thick,dashed] (-4,0) -- (-6,0);
        \draw[<->,thick] (-5,-1.3) -- (-5,1.3) node[midway, above left] {\large$\mathcal{T}_i$};
        \draw[<->,thick,rotate=-90] (-5,-1.3) -- (-5,1.3) node[midway, above left] {\large$\bar{\mathcal{T}}_i$};
        \draw[->,thick] (-28:5.3) arc (-28:28:5.3) node[midway, right] {$\mathbb{Z}_{2k}$};
    \end{tikzpicture}
    \caption{The $2k$-replicated manifold on which the path integral for the replicated fidelity $F_k$ is performed. The arrows show the actions of $\mathbb{Z}_{2k}$ replica symmetry, and the two types of $\mathbb{Z}_2+\mathcal{C}$ symmetry called $\mathcal{T}_i$ and $\bar{\mathcal{T}}_i$. (The case shown is for $k=3$.)}
    \label{Figure: replicated path integral manifold}
\end{figure}

For the states we are considering, one may compute \eqref{Equation: replica fidelity} with a path integral. The manifold over which this path integral should be evaluated contains $2k$ copies of each of $\mathcal{M}^-$ and $\mathcal{M}^+$, which we label $\mathcal{M}^-_i,\mathcal{M}^+_i$ respectively, with $i=1,\dots,2k$ (we will use notation in which the index $i$ is taken mod $2k$). The subregions $A$ and $\bar{A}$ will be labelled $A_i,\bar{A}_i$ respectively in $\partial\mathcal{M}^-_i$, and we will temporarily use $A_i^+,\bar{A}_i^+$ to label their mirror images in $\partial\mathcal{M}^+_i$. In the path integral for \eqref{Equation: replica fidelity}, one glues $\bar{A}_i$ to $\bar{A}_i^+$, and $A_i$ to $A_{i+1}^+$. One then inserts sources $\lambda_1,\lambda_2$ on $\mathcal{M}_i^-$, and $\lambda_1^{\mathrm{T}*},\lambda_2^{\mathrm{T}*}$ on $\mathcal{M}_i^+$, for (w.l.o.g.) odd/even $i$ respectively. In this way one constructs a path integral on a $2k$-fold replicated version of the original manifold, which is portrayed in Figure~\ref{Figure: replicated path integral manifold}.

At large $N$, we may use the holographic dictionary to write the replica path integral in terms of the bulk action. Let $S^{(k)}(\lambda_1,\lambda_2)$ be the gravitational action evaluated on the on-shell bulk field configuration $\phi$ whose boundary conditions are set by the sources in the replica path integral as described above. Then at leading order in $N$ we have
\begin{equation}
    F_k = \exp(kS(\lambda_1,\lambda_1)+kS(\lambda_2,\lambda_2) - S^{(k)}(\lambda_1,\lambda_2)),
    \label{Equation: large N Fk}
\end{equation}
where the contributions of $S(\lambda_1,\lambda_1)$ and $S(\lambda_2,\lambda_2)$ come from the normalisation in \eqref{Equation: rho large N}.

When $\lambda_1=\lambda_2$, the replica path integral has the following symmetries:
\begin{itemize}
    \item One may cyclically permute the individual replicas, sending $\mathcal{M}^-_i\to\mathcal{M}^-_{i+1}$ and $\mathcal{M}^+_i\to\mathcal{M}^+_{i+1}$. This $\mathbb{Z}_{2k}$ symmetry is called replica symmetry.
    \item One may reflect the entire replicated manifold across $A_i \cup A_{i+k}$ (i.e.\ the black circles in Figure~\ref{Figure: replicated path integral manifold}), while also taking the complex conjugate of all the sources. This is a version of the original $\mathbb{Z}_2+\mathcal{C}$ symmetry. We refer to this as $\mathcal{T}_i$ symmetry.
    \item Similarly, one may reflect the entire replicated manifold across $\bar{A}_i \cup \bar{A}_{i+k}$ (i.e.\ the blue circles in Figure~\ref{Figure: replicated path integral manifold}), while also taking the complex conjugate of all the sources. This is another version of $\mathbb{Z}_2+\mathcal{C}$ symmetry, and we refer to it as $\bar{\mathcal{T}}_i$ symmetry.
\end{itemize}
However, when $\lambda_1\ne\lambda_2$, some of these symmetries are violated:
\begin{itemize}
    \item Replica symmetry is broken down to the subgroup $\mathbb{Z}_k$ generated by the permutation sending $\mathcal{M}^-_i\to\mathcal{M}^-_{i+2}$ and $\mathcal{M}^+_i\to\mathcal{M}^+_{i+2}$.
    \item $\mathcal{T}_i$ symmetry is broken entirely.
    \item However, $\bar{\mathcal{T}}_i$ symmetry is maintained.
\end{itemize}
We will assume that the symmetries that hold on the boundary continue to hold in the bulk, i.e.\ they are not spontaneously broken.

From now on, we will write $\lambda_2=\lambda_1+\delta_1\lambda$, and assume that $\delta_1\lambda$ is small. Let us use $\phi_1$ to denote the bulk fields at $\delta_1\lambda=0$. We can decompose this bulk in the following way. Let $\Upsilon$ be the codimension 2 surface in the bulk which is fixed by replica symmetry. Also, let $\Sigma_i,\bar{\Sigma}_i$ be codimension 1 surfaces extending from $\Upsilon$ to $A_i,\bar{A}_i$ respectively, such that $\mathcal{T}_i$ fixes $\Sigma_i$, $\bar{\mathcal{T}}_i$ fixes $\bar{\Sigma}_i$, and replica symmetry maps $\Sigma_i\to\Sigma_{i+1}$, $\bar{\Sigma}_i\to\bar{\Sigma}_{i+1}$. We can then divide the bulk into $4k$ pieces $\mathcal{N}^-_i$ and $\mathcal{N}^+_i$, where $\mathcal{N}^-_i$ is bounded by $\Sigma_i\cup\bar{\Sigma}_i\cup\mathcal{M}_i^-$, and $\mathcal{N}^+_i$ is bounded by $\bar{\Sigma}_i\cup\Sigma_{i+1}\cup\mathcal{M}^+_i$. This is depicted in Figure~\ref{Figure: replica bulk}, which may be thought of as the cross section of Figure~\ref{Figure: replicated path integral manifold} when cut along the plane of the page.
\begin{figure}
    \centering
    \begin{tikzpicture}[thick,scale=1.1]
        \fill[red!20] (0,0) circle (4);
        \foreach \x in {1,2,...,6} {
            \draw[red,dashed,rotate=60*\x+30] (0,0) .. controls (1.5,0.5) and (2.5,-0.5) .. (4,0);
            \fill[red!20] (60*\x+30:1.8) circle (0.3);
            \draw[dashed,rotate=60*\x] (0,0) .. controls (1.5,0.5) and (2.5,-0.5) .. (4,0);
            \fill[red!20] (60*\x:1.8) circle (0.3);
        }
        \foreach \x in {0,1,2} {
            \draw[line width=2pt, blue!80!black] (120*\x:4) arc (120*\x:120*\x+60:4);
            \draw[line width=2pt, blue!60] (120*\x+60:4) arc (120*\x+60:120*\x+120:4);
        }
        \foreach \x in {1,2,...,6} {
            \fill (60*\x:4) circle (0.1);
            \fill[blue!80] (60*\x+30:4) circle (0.1);
            \node at (60*\x-60:4.5) {$A_\x$};
            \node at (60*\x-30:4.5) {$\bar{A}_\x$};
            \node at (60*\x-60:1.8) {$\Sigma_{\x}$};
            \node at (60*\x-30:1.8) {$\bar{\Sigma}_{\x}$};
            \node at (60*\x-48:3) {$\mathcal{N}^-_{\x}$};
            \node at (60*\x-18:3) {$\mathcal{N}^+_{\x}$};
            \node at (60*\x-45:4.5) {$\mathcal{M}^-_{\x}$};
            \node at (60*\x-15:4.5) {$\mathcal{M}^+_{\x}$};
        }
        \fill[red!20] (0,0.4) circle (0.3);
        \fill[red] (0,0) circle (0.1);
        \node at (0,0.4) {$\Upsilon$};
    \end{tikzpicture}
    \caption{The bulk corresponding to $F_k$. Assuming the symmetries of the boundary continue to hold in the bulk, we may decompose it as shown. (The case shown is for $k=3$.)}
    \label{Figure: replica bulk}

    \vspace*{\floatsep}
    \vspace*{0.5in}

    \begin{tikzpicture}[thick]
        \draw[red,dashed,rotate=30,fill=red!20] (0,0) .. controls (1.5,0.5) and (2.5,-0.5) .. (4,0) arc (0:-31:4);
        \draw[red,dashed,rotate=-30,fill=red!20] (0,0) .. controls (1.5,0.5) and (2.5,-0.5) .. (4,0) arc (0:31:4);
        \draw[dashed] (0,0) .. controls (1.5,0.5) and (2.5,-0.5) .. (4,0);
        \draw[line width=2pt, blue!80!black] (0:4) arc (0:30:4);
        \draw[line width=2pt, blue!60] (0:4) arc (0:-30:4);
        \fill (0:4) circle (0.1);
        \fill[blue!80] (30:4) circle (0.1);
        \fill[blue!80] (-30:4) circle (0.1);
        \node at (0:4.5) {$A_1$};
        \node at (30:4.5) {$\bar{A}_1$};
        \node at (-30:4.5) {$\bar{A}_{2k}$};
        \fill[red!20] (0:1.8) circle (0.3);
        \node at (0:1.8) {$\Sigma_1$};
        \node at (40:2.2) {$\bar{\Sigma}_1$};
        \node at (-40:2.2) {$\bar{\Sigma}_{2k}$};
        \node at (12:3) {$\mathcal{N}^-_1$};
        \node at (-18:3) {$\mathcal{N}^+_{2k}$};
        \node at (15:4.5) {$\mathcal{M}^-_1$};
        \node at (-15:4.5) {$\mathcal{M}^+_{2k}$};
        \fill[red] (0,0) circle (0.1);
        \node at (-0.4,0) {$\Upsilon$};
        \draw[->] (50:1) arc (50:320:1);
        \draw[->,shift={(1,0)}] (45:2.5) arc (45:310:2.5);

        \draw[line width=2pt,->] (5.3,0) -- (6.7,0);
        \begin{scope}[shift={(10.5,0)}]
            \fill[red!20] (0,0) circle (2.5);
            \draw[red,dashed] (0,0) -- (2.5,0);
            \draw[dashed] (0,0) -- (-2.5,0);
            \draw[line width=2pt, blue!80!black] (2.5,0) arc (0:-180:2.5);
            \draw[line width=2pt, blue!60] (2.5,0) arc (0:180:2.5);
            \node[above] at (0,0.05) {$\Upsilon$};
            \draw[->] (15:0.7) arc (15:345:0.7);
            \node at (-0.8,-0.8) {$\pi/k$};
            \node at (0,1.7) {$\mathcal{N}^+$};
            \node at (0,-1.7) {$\mathcal{N}^-$};
            \node at (0,2.8) {$\mathcal{M}^+$};
            \node at (0,-2.8) {$\mathcal{M}^-$};
            \node[above] at (1.5,0) {$\bar\Sigma^+$};
            \node[below] at (1.5,-0) {$\bar\Sigma^-$};
            \fill[red!20] (-1.45,0) circle (0.2);
            \node at (-1.45,0) {$\Sigma$};
            \fill (2.5,0) circle (0.1);
            \fill[blue!80] (-2.5,0) circle (0.1);
            \node at (3,0) {$\bar A$};
            \node at (-3,0) {$A$};
            \fill[red] (0,0) circle (0.1);
        \end{scope}
    \end{tikzpicture}
    \caption{$S^{(k)}$ may be understood in terms of the on-shell action for a bulk manifold $\mathcal{N}$ with a conical defect of opening angle $\pi/k$ at $\Upsilon$. This manifold is constructed by identifying $\bar\Sigma^-=\bar{\Sigma}_1$ with $\bar\Sigma^+=\bar{\Sigma}_{2k}$ in $\mathcal{N}^-_1\cup\mathcal{N}^+_{2k}$.}
    \label{Figure: conical manifold}
\end{figure}

When $\delta_1\lambda$ is allowed to become non-zero, the bulk field configuration picks up a perturbation $\delta_{1,2}\phi$ obeying the linearised equations of motion, and consistent with the $\mathbb{Z}_k$ subgroup of replica symmetry, and $\bar{\mathcal{T}}_i$ symmetry. In terms of the Lagrangian density $L=L[\phi_1+\delta_{1,2}\phi]$, we have
\begin{equation}
    \begin{aligned}
        S^{(k)}(\lambda_1,\lambda_1+\delta_1\lambda) &= \sum_{i=1}^{2k}\qty(\int_{\mathcal{N}^-_i} L + \int_{\mathcal{N}_i^+} L) \\
                                                     &= k\qty(\int_{\mathcal{N}_1^-} L + \int_{\mathcal{N}_1^+} L + \int_{\mathcal{N}_{2k}^-}L + \int_{\mathcal{N}_{2k}^+}L) \\
                                                     &= 2k\operatorname{Re}\qty(\int_{\mathcal{N}_1^-}L + \int_{\mathcal{N}_{2k}^+}L),
    \end{aligned}
    \label{Equation: S(k) on N}
\end{equation}
where the second line follows from $\mathbb{Z}_k$ replica symmetry, while the third line follows from $\bar{\mathcal{T}}_1$ and $\bar{\mathcal{T}_{2k}}$ symmetry. Thus, we can understand $S^{(k)}(\lambda_1,\lambda_2)$ purely in terms of the contribution to the action from $\mathcal{N} = \mathcal{N}^-_1\cup\mathcal{N}^+_{2k}$.

At $\delta_1\lambda=0$, by replica symmetry the fields at $\bar{\Sigma}_1$ are equal to those at $\bar{\Sigma}_{2k}$, so we can smoothly identify these two boundaries of $\mathcal{N}$. The result is a bulk manifold with the same boundary conditions as the density matrix $\rho(\lambda_1)$, and which is smooth everywhere except for a conical singularity of opening angle $\pi/k$ at $\Upsilon$. This is demonstrated in Figure~\ref{Figure: conical manifold}. As $k\to\frac12$, this conical singularity goes away, and we are left with just the action evaluated on the on-shell bulk field configuration matching the boundary conditions of $\rho(\lambda_1)$. Thus, upon analytic continuation to $k=\frac12$, we may write
\begin{equation}
    S^{(k)}(\lambda_1,\lambda_1) \to S(\lambda_1,\lambda_1).
\end{equation}

When $\delta\lambda$ is small but non-zero, we can no longer use replica symmetry to compare the fields at $\bar{\Sigma}_1$ and $\bar{\Sigma}_{2k}$. However, we can still treat the perturbation $\delta_{1,2}\phi$ as living on $\mathcal{N}$. Let us simplify the notation by discarding subscripts when referring to parts of this spacetime, so writing $\Sigma = \Sigma_1$, $\bar{\Sigma}=\bar{\Sigma}_1=\bar{\Sigma}_{2k}$, and so on. Note that the fields at $\bar{\Sigma}_1$ differ from those at $\bar{\Sigma}_{2k}$, and so $\delta_{1,2}\phi$ must be discontinuous when crossing $\bar{\Sigma}$. When we need to refer to the field on either side of $\bar\Sigma$, we will use the notation $\bar\Sigma^- = \bar\Sigma_1$ and $\bar\Sigma^+=\bar\Sigma_{2k}$. Besides the discontinuity at $\bar{\Sigma}$, and the conical defect at $\Upsilon$, $\delta_{1,2}\phi$ is otherwise smooth on $\mathcal{N}$.

We need to characterise the discontinuity at $\bar{\Sigma}$ in such a way that permits an easy analytic continuation in $k$. To do so, let $\delta_{1,2}^\curvearrowleft\phi$ be the bulk field variation obtained by acting on $\delta_{1,2}\phi$ once with $\mathbb{Z}_{2k}$ replica symmetry, and let
\begin{align}
    \delta_1\phi &= \delta_{1,2}\phi + \delta_{1,2}^\curvearrowleft\phi, \\
    \tilde\delta_1\phi &= \delta_{1,2}\phi - \delta_{1,2}^\curvearrowleft\phi.
\end{align}
By linearity, $\delta_1\phi,\tilde\delta_1\phi$ are solutions to the linearised equations of motion. Under the action of $\mathbb{Z}_{2k}$ replica symmetry, $\delta_1\phi,\tilde\delta_1\phi$ change by a $\pm$ sign respectively. From this we may deduce the following about $\delta_1\phi,\tilde\delta_1\phi$ restricted to $\mathcal{N}$:
\begin{itemize}
    \item $\delta_1\phi$ is continuous at $\bar{\Sigma}$, while $\tilde\delta_1\phi$ changes sign when crossing $\bar{\Sigma}$. 
    \item The boundary conditions for $\delta_1\phi$ are given by $\delta\lambda^{\mathrm{T}*}$ on $\mathcal{M}^+$ and $\delta\lambda$ on $\mathcal{M}^-$.
    \item The boundary conditions for $\tilde\delta_1\phi$ are given by $\delta\lambda^{\mathrm{T}*}$ on $\mathcal{M}^+$ and $-\delta\lambda$ on $\mathcal{M}^-$.
\end{itemize}
These three conditions, along with $\bar{\mathcal{T}}_i$ symmetry, are sufficient to determine $\delta_1\phi,\tilde\delta_1\phi$ in the entire bulk replicated manifold, and so must be sufficient to determine $\delta_1\phi,\tilde\delta_1\phi$ in $\mathcal{N}$. They are simple to understand for analytically continued $k$, and one may recover $\delta_{1,2}\phi$ from $\delta_{1,2}\phi = \frac12(\delta_1\phi+\tilde\delta_1\phi)$.

When $k\to\frac12$, the conical defect at $\Upsilon$ vanishes. However, $\Upsilon$ remains important, because it is the boundary of $\bar\Sigma$, which is where $\tilde\delta_1\phi$ is discontinuous. For the usual reasons,\footnote{The main reference is~\cite{Lewkowycz:2013nqa}. Very briefly, one may include a term in the action proportional to the area of $\Upsilon$, in order to allow for the conical singularity. In a saddlepoint approximation this area is minimised, and this effect persists in the limit $k\to\frac12$.} at $k=\frac12$, $\Upsilon$ coincides with the surface of minimal area which is homologous to $A$, i.e.\ the HRT surface.

Also, it should be clear that at $k=\frac12$ we may write $\phi_2 = \phi_1+\delta_1\phi$, where $\phi_2$ is the bulk field configuration for the boundary conditions $\lambda_2$ at $\mathcal{M}^-$ and $\lambda_2^{\mathrm{T}*}$ at $\mathcal{M}^+$. This is because the conditions on $\delta_1\phi$ listed above exactly agree with the conditions on $\phi_2-\phi_1$.

Let us briefly comment on the symmetries of $\mathcal{N}$. 
\begin{itemize}
    \item There is a $\mathbb{Z}_2+\mathcal{C}$ symmetry which reflects everything across $\Sigma\cup\bar\Sigma$, including swapping $\bar\Sigma^-$ and $\bar\Sigma^+$, and complex conjugates the fields. This is in some sense inherited from the $\mathcal{T}_i$ symmetry of the replicated spacetime. Under this symmetry, $\phi_1$ and $\delta_1\phi$ are invariant, but $\tilde\delta\phi\to-\tilde\delta\phi$.
    \item There is another type of $\mathbb{Z}_2+\mathcal{C}$ symmetry, distinct from the first, which acts only on the fields at $\bar\Sigma^\pm$. It reflects all components of the fields in time, and complex conjugates them, but it does \emph{not} swap $\bar\Sigma^-$ and $\bar\Sigma^+$. This symmetry is inherited from the $\bar{\mathcal{T}}_i$ symmetry of the replicated spacetime. All of the fields $\phi,\delta_1\phi,\tilde\delta_1\phi$ are invariant under this symmetry. 
\end{itemize}

We may now analytically continue \eqref{Equation: S(k) on N} to $k=\frac12$, and write
\begin{equation}
    S^{(k)}(\lambda_1,\lambda_1+\delta\lambda) = \operatorname{Re}(S[\phi_{1,2}]),
\end{equation}
where $S[\phi_{1,2}]$ denotes the action evaluated for the field configuration $\phi_{1,2} = \phi_1 + \frac12(\delta_1\phi + \tilde\delta_1\phi)$ on $\mathcal{N}$. Therefore, using \eqref{Equation: large N Fk} we may write the fidelity as
\begin{equation}
    \tr(\sqrt{\sqrt{\rho_1}\rho_2\sqrt{\rho_1}}) = F_{\frac12} = \exp(\frac12 S(\lambda_1,\lambda_1) + \frac12 S(\lambda_2,\lambda_2) - \operatorname{Re}(S[\phi_{1,2}])).
    \label{Equation: holographic fidelity}
\end{equation}
Thus, we have found an expression for the fidelity of the two holographic states $\rho_1,\rho_2$ in terms of the action for the bulk field configuration $\phi_{1,2}$. 

\subsection{Parallel purifications from Uhlmann's theorem}
\label{Section: holographic parallel purifications}

It is the objective of this section to construct parallel purifications $\ket*{\psi_1},\ket*{\psi_2}$ of the two density matrices $\rho(\lambda_1),\rho(\lambda_2)$. This will be done by using the expression for the holographic fidelity \eqref{Equation: holographic fidelity} in Uhlmann's theorem \eqref{Equation: Uhlmann's theorem}.

For the first state we will simply take the normalised version of the state $\ket*{\lambda_1}$ constructed with \eqref{Equation: holographic state with sources}:
\begin{equation}
    \ket*{\psi_1} = e^{\frac12 S(\lambda_1,\lambda_1)}\ket*{\lambda_1}.
    \label{Equation: psi1}
\end{equation}
This is a purification of $\rho(\lambda_1)$ by construction.

\begin{figure}
    \centering
    \begin{subfigure}{0.35\textwidth}
        \centering
        \begin{tikzpicture}
            \draw[blue,fill=red!20,line width=2pt] (0,0) arc (-180:0:2);
            \draw[red,dashed,thick] (0,0) -- (4,0);
            \node[below] at (2,0) {$\varphi$};
            \node[above] at (2,0) {$B$};
            \node[above] at (2,-2) {$\lambda$};
            \node[below] at (2,-2) {$\mathcal{M}^-$};
        \end{tikzpicture}
        \caption{}
        \label{Figure: bulk wavefunctional ket}
    \end{subfigure}
    \begin{subfigure}{0.35\textwidth}
        \centering
        \begin{tikzpicture}
            \draw[blue,fill=red!20,line width=2pt] (0,0) arc (180:0:2);
            \draw[red,dashed,thick] (0,0) -- (4,0);
            \node[above] at (2,0) {$\varphi$};
            \node[below] at (2,0) {$B$};
            \node[below] at (2,2) {$\lambda^{\mathrm{T}*}$};
            \node[above] at (2,2) {$\mathcal{M}^+$};
        \end{tikzpicture}
        \caption{}
        \label{Figure: bulk wavefunctional bra}
    \end{subfigure}
    \caption{The boundary conditions for the bulk wavefunctionals \mbox{\textbf{(a)}\ $\braket*{\varphi}{\lambda}$} and \mbox{\textbf{(b)}\ $\braket*{\lambda}{\varphi}$.}}
    \label{Figure: bulk wavefunctional}
\end{figure}
We will use the bulk theory to construct $\ket*{\psi_2}$, so let us start by recalling some facts. At large $N$, the wavefunctional for the bulk fields in the state $\ket*{\lambda}$ may be computed with a bulk path integral. In particular, one may write
\begin{equation}
    \braket*{\varphi}{\lambda} = \int_\lambda^\varphi \Dd{\phi} e^{-S^-[\phi]},
    \label{Equation: bulk wavefunctional}
\end{equation}
where the integral is done over all bulk field configurations whose boundary conditions are set by $\lambda$ at the asymptotic boundary $\mathcal{M}^-$, and $\varphi$ on a bulk surface $B$, as shown in Figure~\ref{Figure: bulk wavefunctional ket}. We should emphasise our notation here: $\phi$ denotes the bulk fields, whereas $\varphi$ denotes the boundary data at $B$. $S^-[\phi]$ is the bulk action of the field configuration with these boundary conditions. 

The wavefunctional of the dual states $\bra*{\lambda}$ may be written
\begin{equation}
    \braket*{\lambda}{\varphi} = \int_\varphi^\lambda \Dd{\phi} e^{-S^+[\phi]}
    \label{Equation: bulk dual wavefunctional}
\end{equation}
where the integral is done over all bulk field configurations with boundary data $\lambda^{\mathrm{T}*}$ at $\mathcal{M}^+$, and $\varphi$ at $B$. This is shown in Figure~\ref{Figure: bulk wavefunctional bra}. $S^+[\phi]$ is the bulk action for these field configurations.

The states $\ket{\varphi}$ make up a normalised basis for the bulk Hilbert space. The identity in this basis is
\begin{equation}
    \int \Dd{\varphi} \ket{\varphi}\bra{\varphi}.
    \label{Equation: bulk Hilbert space identity}
\end{equation}
This implies that
\begin{equation}
    \begin{aligned}
        \braket{\lambda_2}{\lambda_1} &= \int \Dd{\varphi} \braket{\lambda_2}{\varphi} \braket{\varphi}{\lambda_1} \\
                                      &= \int\Dd{\varphi} \qty(\int_\varphi^{\lambda_2} \Dd{\phi_+} e^{-S_+[\phi_+]})\qty(\int^\varphi_{\lambda_1} \Dd{\phi_-} e^{-S^-[\phi_-]}) \\
                                      &= \int_{\lambda_1}^{\lambda_2} \Dd{\phi} e^{-S[\phi]}.
    \end{aligned}
\end{equation}
Where the last integral is done over all bulk field configurations whose boundary conditions are given by $\lambda_1$ at $\mathcal{M}^-$, and $\lambda_2^{\mathrm{T}*}$ at $\mathcal{M}^+$. The action in the last line is $S[\phi] = S^+[\phi_+] + S^-[\phi_-]$. At large $N$, a stationary phase approximation recovers \eqref{Equation: large N inner product}.

One may determine what kind of boundary conditions are set by the state $\ket{\varphi}$ by looking at the variation of the bulk action $S^-=\int_{\text{bulk}} L$. When the bulk equations of motion are obeyed, one has
\begin{equation}
    \delta S^- = \int_{\mathcal{M}^-}\theta[\phi,\delta\phi] + \int_B\theta[\phi,\delta\phi],
\end{equation}
The contribution at $B$ determines the form of $\varphi$ by the requirement that it vanishes when $\varphi$ is kept fixed. In particular this means that $\int_B\theta$ can only depend on $\delta\phi$ through $\delta\varphi$. For the toy example of a scalar field with $L=-\frac12\dd{\phi}\wedge*\dd{\phi}$, we have $\theta=-\delta\phi *\dd{\phi}$, so $\varphi$ is just the initial data for the scalar field on $B$.

There is an easy way to carry out a change of basis in the bulk Hilbert space: one simply adds a boundary term of the form
\begin{equation}
    S_B = \int_B D[\phi]
\end{equation}
to the action in \eqref{Equation: bulk wavefunctional}. One must simultaneously subtract $S_B$ from the action in \eqref{Equation: bulk dual wavefunctional}. The variation of the new action $S^- = \int_{\text{bulk}} L + \int_B D$ reads
\begin{equation}
    \delta S^- = \int_{\mathcal{M}^-}\theta[\phi,\delta\phi] + \int_B\Big(\theta[\phi,\delta\phi] + \delta D[\phi]\Big).
\end{equation}
This modifies the way in which the term at $B$ depends on $\delta\phi$, and therefore changes the type of boundary data $\varphi$ specified by the bulk state $\ket{\varphi}$. For the example of the scalar field, one might pick $D = \phi *\dd{\phi}$. Then we would have $\theta+\delta D = \phi\,\delta(*\dd{\phi})$, and hence $\varphi$ would be the normal derivative of the scalar field at $B$, i.e.\ its conjugate momentum in a canonical treatment.

We should note that the boundary data $\varphi$ must be invariant under $\mathbb{Z}_2+\mathcal{C}$. This is to ensure that a Wick rotation to a real Lorentzian spacetime exists, and states without this property are not part of the bulk Hilbert space. It is also important that this does \emph{not} mean that the dominant field configuration in any stationary phase approximation must be $\mathbb{Z}_2+\mathcal{C}$ invariant at $B$. This is because stationary phase methods involve a complex deformation of the field contour. For similar reasons, $S_B$ must be imaginary for all $\mathbb{Z}_2+\mathcal{C}$ invariant field configurations. This ensures, for example, that the form of the identity \eqref{Equation: bulk Hilbert space identity} is the same for different possible choices of $S_B$.

Let us identify $B$ with $\Sigma\cup\bar{\Sigma}$, as defined in Section~\ref{Section: fidelity replica}. Let $\phi_1$ be the bulk on-shell field configuration whose boundary conditions match those of $\rho(\lambda_1)$, and let $\delta_1\phi,\tilde\delta_1\phi$ be the field variations such that $\phi_2=\phi_1+\delta_1\phi$, and $\phi_{1,2} = \phi_1+\frac12(\delta_1\phi+\tilde\delta\phi)$ is the configuration relevant to the fidelity of $\rho(\lambda_1),\rho(\lambda_2)$, as discussed in the previous subsection.
\begin{figure}
    \centering
    \begin{tikzpicture}[thick,xscale=-0.6,yscale=0.6,font=\large]
        \path[name path = outercircle] (0,0) circle (4);
        \path[name path = innercircle] (0,0) circle (0.3);
        \path[name path = sigmabarplus] (0,0.1) -- (-4,0.1);
        \path[name path = sigmabarminus] (0,-0.1) -- (-4,-0.1);
        \draw[name intersections={of = outercircle and sigmabarplus}] (intersection-1) coordinate (A);
        \draw[name intersections={of = innercircle and sigmabarplus}] (intersection-1) coordinate (B);
        \draw[name intersections={of = outercircle and sigmabarminus}] (intersection-1)  coordinate (C);
        \draw[name intersections={of = innercircle and sigmabarminus}] (intersection-1)  coordinate (D);
        \coordinate (O) at (0,0);
        \pgfextractangle{\angleA}{O}{A}
        \pgfextractangle{\angleB}{O}{B}
        \pgfextractangle{\angleC}{O}{C}
        \pgfextractangle{\angleD}{O}{D}
        \fill[red!20] (A) -- (B) arc (\angleB:\angleD-360:0.3) -- (C) arc (\angleC-360:\angleA:4);
        \draw[red,line width=1.5pt] (A) -- (B) arc (\angleB:\angleD-360:0.3) -- (C);
        \draw[blue,line width=2pt] (C) arc (\angleC-360:\angleA:4);
        \draw[dashed] (0.3,0) -- (4,0);
        \node[above] at (-2,0.1) {$\overline{\Sigma}^+$};
        \node[below] at (-2,-0.1) {$\overline{\Sigma}^-$};
        \fill[red!20] (2,0) circle (0.3);
        \node at (2,0) {$\Sigma$};
        \fill[red] (0,0) circle (0.1);
        \node[above] at (0,0.3) {$B_\Upsilon$};
        \node[below] at (0,-4) {$\mathcal{M}^-$};
        \node[above] at (0,4) {$\mathcal{M}^+$};
    \end{tikzpicture}
    \caption{Variations of the action in the presence of discontinuities at $\bar\Sigma$ lead to new boundary terms at $\bar\Sigma^-\cup\bar\Sigma^+\cup B_\Upsilon$.}
    \label{Figure: bulk without bar sigma}
\end{figure}
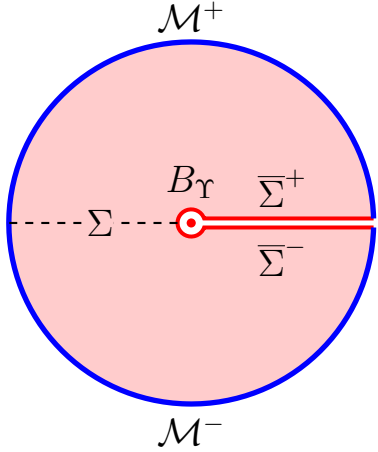

Consider the classical phase space for the bulk field theory. We can view points in this phase space as describing the fields at $B$. Consider a map $Z_{1,2}$ from phase space to itself, whose action does not affect the fields at $\Sigma$, but maps $\phi_{1,2}|_{\bar\Sigma^-}$ to $\phi_{1,2}|_{\bar\Sigma^+}$ at $\bar\Sigma$ (these differ by $\tilde\delta_1\phi|_{\bar\Sigma^-}$). This defines the action of $Z_{1,2}$ at certain points in phase space, and we may extend $Z_{1,2}$ to a symplectomorphism of the full classical phase space (a.k.a.\ a canonical transformation of the classical variables). In the quantum theory there is a corresponding unitary operator $X_{1,2}$ which implements the action of $Z_{1,2}$. Let $\varphi_{1,2}^\pm$ be the boundary data for $\phi_{1,2}$ on $\Sigma\cup\bar\Sigma^\pm$ respectively. It is clear that
\begin{equation}
    X_{1,2}\ket*{\varphi_{1,2}^-} = e^{ix_{1,2}}\ket*{\varphi_{1,2}^+},
    \label{Equation: X varphi}
\end{equation}
where $x_{1,2}$ is some real number that we will leave undetermined.

Let us see what happens when we insert this operator in between $\bra{\lambda_2}$ and $\ket{\lambda_1}$. Using path integral notation, we have
\begin{equation}
    \mel{\lambda_2}{X_{1,2}}{\lambda_1} = \int^{\lambda_2}_{\lambda_1} \Dd{\phi_+}\Dd{\phi_-} e^{-S^+[\phi_+]} X_{1,2} e^{-S^-[\phi_-]}.
\end{equation}
The presence of $X_{1,2}$ means that the integral is done over field configurations with the property that the fields $\phi_+$ at $B$ are related to the fields $\phi_-$ at $B$ by an action of $Z_{1,2}$. Also, the boundary conditions $\lambda_1,\lambda_2^{\mathrm{T}*}$ at $\mathcal{M}^-,\mathcal{M}^+$ must continue to hold.

An important question to ask is whether we can still use a stationary phase approximation to evaluate this integral. A variation of the field configurations $\phi_1,\phi_2\to\phi_1+\delta\phi_2,\phi_2+\delta\phi_2$ in the integrand leads to an expression of the form
\begin{equation}
    e^{-S[\phi_2] + \int E[\phi_2]\cdot\delta\phi_2} e^{-\int_B \theta[\phi_2,\delta\phi_2]} X_{1,2} e^{\int_B \theta[\phi_1,\delta\phi_1]} e^{-S[\phi_1] + \int E[\phi_1]\cdot\delta\phi_1}.
    \label{Equation: X12 integrand variation}
\end{equation}
So in addition to the equations of motion, we pick up some terms at $B$, which could potentially be an issue. However, we should view $e^{-\int_B \theta[\phi_2,\delta\phi_2]}$ as a bulk operator, and when commuted past $X_{1,2}$, the fields in this operator are acted upon by $Z_{1,2}$. Thus,
\begin{equation}
    e^{-\int_B \theta[\phi_2,\delta\phi_2]} X_{1,2} e^{\int_B \theta[\phi_1,\delta\phi_1]} = X_{1,2} e^{-\int_B \theta[\phi_1,\delta\phi_1]} e^{\int_B \theta[\phi_1,\delta\phi_1]} = X_{1,2},
\end{equation}
and so \eqref{Equation: X12 integrand variation} becomes
\begin{equation}
    e^{-S[\phi_2] + \int E[\phi_2]\cdot\delta\phi_2} X_{1,2} e^{-S[\phi_1] + \int E[\phi_1]\cdot\delta\phi_1},
\end{equation}
i.e.\ only the equations of motion remain. Thus, the stationary phase method still works. In particular, the path integral is dominated by field configurations which pick up an action of $Z_{1,2}$ when crossing $B$, and which obey the equations of motion elsewhere.

It seems like the dominant field configuration should be exactly $\phi_{1,2}$. However, there is one additional constraint that must be satisfied. Consider the change in the bulk action under an on-shell variation of the bulk fields $\phi_{1,2}\to\phi_{1,2} + \delta\phi$. Because of the discontinuities at $\bar\Sigma$, we should evaluate this action on a spacetime with $\bar\Sigma$ removed, and this introduces new boundaries $\bar\Sigma^-\cup\bar\Sigma^+\cup B_\Upsilon$, as shown in Figure~\ref{Figure: bulk without bar sigma}. The surfaces $\bar\Sigma^-,\bar\Sigma^+$ are just on either side of $\bar\Sigma$, while $B_\Upsilon$ wraps around $\Upsilon$. Assuming that the holographic boundary conditions $\lambda_1,\lambda_2$ are fixed, the variation of the action reduces to the boundary terms
\begin{equation}
    \delta S[\phi_{1,2}] = \int_{\bar\Sigma^-}\theta[\phi_{1,2},\delta\phi] - \int_{\bar\Sigma^+} \theta[\phi_{1,2},\delta\phi] + \int_{B_\Upsilon}\theta[\phi_{1,2},\delta\phi].
    \label{Equation: delta S phi 12}
\end{equation}
The terms at $\bar\Sigma^-,\bar\Sigma^+$ do not concern us because the presence of $X_{1,2}$ means they cancel in the path integral, as just explained. However, the contribution at $B_\Upsilon$ is in general non-vanishing, and interacts with $X_{1,2}$ in a complicated way. 

Recall from the Introduction that there is an ambiguity in the definition of $\theta$. In particular, we are free to carry out the change $\theta[\phi,\delta\phi]\to\theta[\phi,\delta\phi]+\dd{K[\phi,\delta\phi]}$. We will use this change to remove the term at $B_\Upsilon$ in \eqref{Equation: delta S phi 12}. In other words, we pick a $K$ such that
\begin{equation}
    \int_{B_\Upsilon}\theta[\phi_{1,2},\delta\phi] = - \int_{\partial B_\Upsilon}K[\phi_{1,2},\delta\phi].
\end{equation}
Note that each of these integrals should be considered in a limit in which $B_\Upsilon$ tightly encloses $\Upsilon$, so the left-hand side really only depends on the fields at $\Upsilon$, and the right-hand side becomes an integral at $\Upsilon$. Thus, it is possible to choose a $K[\phi,\delta\phi]$ satisfying the above in such a way that it only depends on the field configuration locally. From now on we will assume that we have done this transformation $\theta\to\theta+\dd{K}$, so that the boundary term
\begin{equation}
    \int_{B_\Upsilon} \theta[\phi_{1,2},\delta\phi]
\end{equation}
vanishes. This will be crucial for the resolution of the boundary ambiguity in the covariant phase space formalism, and will be discussed further in Section~\ref{Section: ambiguity resolution}.

Note that $X_{1,2}$ commutes with observables on $\Sigma$. Bulk reconstruction~\cite{Almheiri:2014lwa,Dong:2016eik} implies that $X_{1,2}$ can therefore be treated as a unitary operator in the boundary theory acting on $\mathcal{H}_{\bar{A}}$. By this we mean that one may write $X_{1,2}=I_A\otimes X_{\bar{A}}$ with respect to the decomposition $\mathcal{H}=\mathcal{H}_A\otimes\mathcal{H}_{\bar{A}}$, where $I_A$ is the identity acting on $\mathcal{H}_A$, and $X_{\bar{A}}$ is a unitary operator acting on $\mathcal{H}_{\bar{A}}$.

We are now ready to construct the state $\ket*{\psi_2}$. It is given by
\begin{equation}
    \ket*{\psi_2} = e^{\frac12 S(\lambda_2,\lambda_2)} X_{1,2}^\dagger \ket*{\lambda_2}.
    \label{Equation: psi2}
\end{equation}
Since $X_{1,2}$ acts on $\mathcal{H}_{\bar{A}}$, this is a genuine purification of $\rho(\lambda_2)$. It remains to show that $\ket*{\psi_1}$ and $\ket*{\psi_2}$ are parallel. To do this, note that the stationary phase approximation allows us to write
\begin{equation}
    \mel*{\lambda_2}{X_{1,2}}{\lambda_1} = e^{-S[\phi_{1,2}]}\mel*{\varphi_{1,2}^+}{X_{1,2}}{\varphi_{1,2}^-}.
    \label{Equation: X12 scattering}
\end{equation}
Using \eqref{Equation: X varphi}, the second factor on the right-hand side is equal to $e^{ix_{1,2}}$. Including the normalising factors in \eqref{Equation: psi1} and \eqref{Equation: psi2}, and taking the absolute value, one finds
\begin{equation}
    |\braket*{\psi_2}{\psi_1}| = \exp(\frac12 S(\lambda_1,\lambda_1) + \frac12 S(\lambda_2,\lambda_2) - \Re S[\phi_{1,2}]).
\end{equation}
This matches exactly with the holographic fidelity found in the previous section. Therefore, by Uhlmann's theorem, $\ket*{\psi_1}$ and $\ket*{\psi_2}$ are parallel.

\subsection{Uhlmann phase}
\label{Section: holographic Uhlmann phase}
By using the results of the previous subsection, we will now compute the Uhlmann phase associated with a closed curve of holographic density matrices $\rho(\lambda)$ reduced to a fixed subregion $A$ arising from a closed curve $\lambda(t)$ of boundary sources.

First let us state our notation. Let $\lambda_i$, $i=1,\dots,n$ be a sequence of points ordered along the curve of sources, and let $\rho_i = \rho(\lambda_i)$ be the density matrices for these sources, obtained by reducing the pure states $\ket{\lambda_i}$ (as prepared using \eqref{Equation: holographic state with sources}) to $\mathcal{H}_A$:
\begin{equation}
    \rho_i = \tr_{\bar{A}}\ket{\lambda_i}\bra{\lambda_i}.
\end{equation}
Let $\phi_i$ be the bulk field configuration matching the boundary conditions set by the boundary sources $\lambda_i$, and let $\delta_i\phi$ be defined by $\phi_{i+1}=\phi_i+\delta_i\phi$. Furthermore, let $\phi_{i,i+1} = \phi_i+\frac12(\delta_i\phi+\tilde\delta_i\phi)$ be the bulk field configuration relevant to the fidelity of $\rho(\lambda_i)$ and $\rho(\lambda_{i+1})$. We may construct symplectomorphisms $Z_{i,i+1}$ of the bulk phase space that map $\phi_{i,i+1}|_{\bar\Sigma^-}$ to $\phi_{i,i+1}|_{\bar\Sigma^+}$, and associated unitary operators $X_{i,i+1}$. Let us define the real numbers $x_{i,i+1}$ with
\begin{equation}
    X_{i,i+1}\ket{\varphi_{i,i+1}^-} = e^{ix_{i,i+1}}\ket{\varphi_{i,i+1}^+},
\end{equation}
where $\varphi_{i,i+1}^\pm$ is the boundary data for $\phi_{i,i+1}$ at $\Sigma\cup\bar\Sigma^\pm$ respectively.

Consider the following sequence of states:
\begin{align}
    \ket{\psi_1} &= e^{\frac12 S(\lambda_1,\lambda_1)} \ket{\lambda_1}, \\
                 &\dots \nonumber\\
    \ket{\psi_i} &= e^{\frac12 S(\lambda_i,\lambda_i)} X_{1,2}^\dagger X_{2,3}^\dagger \dots X^\dagger_{i-1,i} \ket{\lambda_i} \\
                 &\dots \nonumber\\
    \ket{\psi_n} &= e^{\frac12 S(\lambda_n,\lambda_n)} X_{1,2}^\dagger X_{2,3}^\dagger \dots X^\dagger_{n-1,n} \ket{\lambda_n}.
\end{align}
Since each $X_{i,i+1}$ is a unitary operator acting on $\mathcal{H}_{\bar{A}}$, it is clear that $\ket{\psi_i}$ is a purification of $\ket{\lambda_i}$ for all $i$. The prefactors involving $S(\lambda_i,\lambda_i)$ ensure these states are normalised. Furthermore, we have
\begin{align}
    \braket{\psi_{i+1}}{\psi_i} &= e^{\frac12 S(\lambda_i,\lambda_i)}e^{\frac12 S(\lambda_{i+1},\lambda_{i+1})} \mel*{\lambda_{i+1}}{X_{i,i+1}}{\lambda_i}\\
                                &= \exp(\frac12 S(\lambda_i,\lambda_i) + \frac12 S(\lambda_{i+1},\lambda_{i+1}) - S[\phi_{i,i+1}] + ix_{i,i+1}).
    \label{Equation: psi overlaps}
\end{align}
where the second line follows from the same logic as in \eqref{Equation: X12 scattering}, and we are assuming that
\begin{equation}
    \int_{B_\Upsilon} \theta[\phi_{i,i+1},\delta\phi] = 0.
    \label{Equation: theta condition}
\end{equation}
The condition \eqref{Equation: theta condition} may be enforced by a suitable redefinition $\theta\to\theta+\dd{K}$, as described previously. In order for the bulk Hilbert space basis $\ket*{\varphi}$ to be the same throughout the proceeding argument, this redefinition must be done using a $K$ that is the same for all $i$. This is possible, and in fact one may enforce the stronger condition 
\begin{equation}
    \int_{B_\Upsilon} \theta[\phi,\delta\phi] = 0,
    \label{Equation: theta condition stronger}
\end{equation}
where here $\phi,\delta\phi$ can be any field configuration and variation that might be discontinuous at $\bar\Sigma$, but obey the equations of motion elsewhere. Indeed, if \eqref{Equation: theta condition stronger} is not true, then we may pick a $K$ such that
\begin{equation}
    \int_{B_\Upsilon} \theta[\phi,\delta\phi] = -\int_{\partial B_\Upsilon} K[\phi,\delta\phi].
\end{equation}
In the limit as $B_\Upsilon$ tightly encloses $\Upsilon$, both sides only depend on the fields at $\Upsilon$. Thus we may choose $K$ such that it only has a local dependence on the fields. After redefining $\theta\to\theta+\dd{K}$, one then gets a $\theta$ satisfying \eqref{Equation: theta condition stronger}.

One notes that $\abs*{\braket*{\psi_{i+1}}{\psi_i}}$ matches with the fidelity of $\rho_i,\rho_{i+1}$. Thus, each consecutive pair $\ket{\psi_i},\ket{\psi_{i+1}}$ of the above states is parallel. One therefore can obtain the Uhlmann phase by computing the quantity
\begin{equation}
    e^{i\gamma} = \lim_{n\to\infty} \braket*{\psi_1}{\psi_n}\braket*{\psi_n}{\psi_{n-1}}\dots\braket*{\psi_2}{\psi_1}.
    \label{Equation: ye uhlmann}
\end{equation}
All but the first factor in this object may be computed using \eqref{Equation: psi overlaps}. It thus remains to compute  
\begin{equation}
    \braket*{\psi_1}{\psi_n} = e^{\frac12S(\lambda_1,\lambda_1)}e^{\frac12S(\lambda_n,\lambda_n)} \mel*{\lambda_1}{X^\dagger}{\lambda_n},
\end{equation}
where
\begin{equation}
    X^\dagger = X_{1,2}^\dagger X_{2,3}^\dagger \dots X^\dagger_{n-1,n}.
\end{equation}

Recall from the previous section that one may carry out a change of basis for the bulk Hilbert space by modifying the action by a boundary term $S_B=\int_B D$. It will be convenient for us to choose a $D[\phi]$ obeying the following condition for all $i$:
\begin{equation}
    (\delta_i+\tilde\delta_i)\int_{\bar\Sigma^-}D[\phi_i] = -\int_{\bar\Sigma^-}\theta[\phi_i,\delta_i\phi+\tilde\delta_i\phi].
    \label{Equation: boundary D general}
\end{equation}
If we view $\int_{\bar\Sigma^-}D$ as a function on field space, it is clear that we can satisfy this condition, because all we need to do is ensure that the derivative of this function at $\phi_i$ in the direction $\delta_i\phi+\tilde\delta_i\phi$ is given by the right-hand side. As mentioned previously, all the fields $\phi_i, \delta_i\phi, \tilde\delta_i\phi$ are invariant under $\mathbb{Z}_2+\mathcal{C}$ symmetry.\footnote{To be clear, here we are considering the action of $\mathbb{Z}_2$ in such a way that it does \emph{not} swap $\bar\Sigma^-$ and $\bar\Sigma^+$ -- it merely applies a time reflection to all the components of the fields at $\bar\Sigma^-$.} Therefore, the right-hand side of \eqref{Equation: boundary D general} is imaginary (since $\mathbb{Z}_2$ changes the orientation of $\bar\Sigma^-$). Thus, this condition is consistent with the requirement that $S_B$ be imaginary. Also, it is clear that this condition can be satisfied by a regular function $S_B$ even as $n\to\infty$, because the vector $\delta\phi+\tilde\delta\phi$ is never parallel to the curve of field configurations $\phi$. It may not be possible to choose $D$ such that \eqref{Equation: boundary D general} is obeyed for \emph{all} possible curves of field configurations, but this will end up not being important for our final result.

It is in principle unnecessary to enforce \eqref{Equation: boundary D general}, and not doing so should still lead to the same results obtained in this paper. However, we are free to enforce it, and this will be useful in what follows. For notational convenience, we will absorb $\delta D$ into the definition of $\theta$. Having done so, \eqref{Equation: boundary D general} means that we can assume
\begin{equation}
    \int_{\bar\Sigma^-}\theta[\phi_i,\delta_i\phi+\tilde\delta_i\phi] = 0.
    \label{Equation: redefined theta}
\end{equation}
This then implies a certain choice of basis $\ket{\varphi}$ for the bulk Hilbert space.

Note that
\begin{equation}
    \begin{aligned}
        \varphi_{i,i+1}^+ &= \text{boundary data for } \phi_i + \textstyle\frac12(\delta_i\phi+\tilde\delta_i\phi) \text{ at } \bar\Sigma^+ \\
                          &= \text{boundary data for } \phi_i + \textstyle\frac12(\delta_i\phi-\tilde\delta_i\phi) \text{ at } \bar\Sigma^- \\
                          &= \text{boundary data for } \phi_i + \delta_i\phi \text{ at } \bar\Sigma^- \\
                          &= \text{boundary data for } \phi_{i+1} \text{ at } \bar\Sigma^- \\
                          &= \text{boundary data for } \phi_{i+1} + \textstyle\frac12(\delta_{i+1}\phi+\tilde\delta_{i+1}\phi) \text{ at } \bar\Sigma^- \\
                          &= \varphi^-_{i+1,i+2},
    \end{aligned}
    \label{Equation: phi* i = phi i+1}
\end{equation}
where the third and fifth lines follow from \eqref{Equation: redefined theta}. Therefore the two states $\ket*{\varphi_{i,i+1}^+}$ and $\ket*{\varphi_{i+1,i+2}^-}$ are actually equivalent. Thus, using 
\begin{equation}
    X_{i,i+1}^\dagger \ket*{\varphi_{i+1,i+2}^-} = X_{i,i+1}^\dagger\ket*{\varphi_{i,i+1}^+} = e^{-ix_{i,i+1}}\ket*{\varphi_{i,i+1}^-},
\end{equation}
we have
\begin{equation}
    X^\dagger\ket*{\varphi_{n,1}^-} = \exp(-i\sum_{i=1}^{n-1}x_{i,i+1})\ket*{\varphi_{1,2}^-} = \exp(-i\sum_{i=1}^{n-1}x_{i,i+1})\ket*{\varphi^+_{n,1}}.
    \label{Equation: X dagger action}
\end{equation}
Here we are defining $\varphi_{n,1}^\pm$ as the boundary data at $\Sigma\cup\bar{\Sigma}^\pm$ respectively for the field configuration $\phi_{n,1}+\frac12(\delta_n\phi+\tilde\delta_n\phi)$, which is the one relevant to the fidelity of $\rho_n,\rho_1$. Because the curve of density matrices is closed, \eqref{Equation: phi* i = phi i+1} applies for $\varphi_{n,1}^-,\varphi_{n,1}^+$ also, if we treat $i$ as an index mod $n$.

The operator $X^\dagger$ corresponds to the symplectomorphism $Z_{1,2}^{-1}Z_{2,3}^{-1}\dots Z_{n-1,n}^{-1}$. At leading order in the limit $n\to\infty$, the operator $Z_{i,i+1}^{-1}$ can be treated as carrying out the infinitesimal change $\phi\to\phi-\tilde\delta_i\phi|_{\bar\Sigma^-}$. Thus $Z_{1,2}^{-1}Z_{2,3}^{-1}\dots Z_{n-1,n}^{-1}$ approximately acts as (in an appropriately linearised sense)
\begin{equation}
    \phi\to\phi-\sum_{i=1}^{n-1}\tilde\delta_i\phi|_{\bar\Sigma^-}.
\end{equation}
But note that
\begin{equation}
    -\sum_{i=1}^{n-1}\tilde\delta_i\phi \approx \tilde\delta_n\phi.
    \label{Equation: delta n - phi}
\end{equation}
This can be understood by considering the change in boundary conditions at asymptotic infinity for each $\tilde\delta_i\phi$. The fact that the curve of boundary conditions is closed implies that
\begin{equation}
    \sum_{i=1}^n \delta_i\lambda \approx 0,
\end{equation}
from which one obtains \eqref{Equation: delta n - phi}. Thus, at leading order in the limit $n\to\infty$, the operator $X^\dagger$ corresponds to a symplectomorphism $Z^{-1}$ carrying out the change $\phi\to\phi+\tilde\delta_n\phi|_{\bar\Sigma}$. This maps $\phi_{n,1}|_{\bar{\Sigma}^-}$ to $\phi_{n,1}|_{\bar\Sigma^+}$. 

Consider now the correlator
\begin{equation}
    \mel{\lambda_1}{X^\dagger}{\lambda_n} = \int \Dd{\phi_+}\Dd{\phi_-} e^{-S^+[\phi_+]} X^\dagger e^{-S^-[\phi_-]}.
\end{equation}
We sum over fields obeying the boundary conditions $\lambda_n,\lambda_1^{\mathrm{T}*}$ at $\mathcal{M}^-,\mathcal{M}^+$. By the same logic as in the previous section, we can use a stationary phase approximation to compute this integral. It is dominated by the field configuration which picks up an action of $Z^{-1}$ when crossing $B$, and which obey the equations of motion elsewhere. This field configuration is $\phi_{n,1}$. Thus we have
\begin{equation}
    \mel{\lambda_1}{X^\dagger}{\lambda_n} = e^{-S[\phi_{n,1}]} \mel*{\varphi_{n,1}^+}{X^\dagger}{\varphi_{n,1}^-}.
\end{equation}
By \eqref{Equation: X dagger action}, the latter factor on the right-hand side is $\exp(-i\sum_{i=1}^{n-1}x_{i,i+1})$.

Using this, we can compute the final scattering amplitude in the Uhlmann phase. In fact, with the choices we have made, it conveniently takes a form similar to all the other factors. It is
\begin{equation}
    \braket{\psi_1}{\psi_n} = \exp(\frac12S(\lambda_1,\lambda_1) + \frac12S(\lambda_n,\lambda_n) - S[\phi_{n,1}] - i\sum_{i=1}^{n-1}x_{i,i+1}).
\end{equation}
In total, the Uhlmann phase may be written
\begin{equation}
    e^{i\gamma} = \lim_{n\to\infty} \prod_{i=1}^n \exp(\frac12 S(\lambda_i,\lambda_i) + \frac12 S(\lambda_{i+1},\lambda_{i+1}) - S[\phi_{i,i+1}]),
    \label{Equation: Uhlmann phase 1}
\end{equation}
where all the terms involving the numbers $x_{i,i+1}$ exactly cancel.

Let us compute the contribution of each term. Using
\begin{align}
    S(\lambda_i,\lambda_i) &= S[\phi_i], \\
    S(\lambda_{i+1},\lambda_{i+1}) &= S[\phi_i] + \delta_i S[\phi_i] + \dots, \\
    S[\phi_{i,i+1}] &= S[\phi_i] + \frac12(\delta_i+\tilde\delta_i) S[\phi_i] + \dots,
\end{align}
one finds
\begin{equation}
    \frac12 S(\lambda_i,\lambda_i)+\frac12 S(\lambda_{i+1},\lambda_{i+1}) - S[\phi_{i,i+1}] = -\frac12\tilde\delta_i S[\phi_i].
\end{equation}
In terms of boundary integrals, one has
\begin{equation}
    -\frac12\tilde\delta_i S[\phi_i] = -\frac12\int_{\mathcal{M}^-\cup\mathcal{M}^+\cup\bar\Sigma^-\cup\bar\Sigma^+}\theta[\phi_i,\tilde\delta_i\phi] = -\int_{\mathcal{M}^-\cup\bar\Sigma^-} \theta[\phi_i,\tilde\delta_i\phi].
    \label{Equation: tilde delta S}
\end{equation}
Note that, at leading order in the variations, using \eqref{Equation: theta condition} we may write
\begin{equation}
    \int_{B_\Upsilon}\theta[\phi_i,\tilde\delta_i\phi] = \int_{B_\Upsilon}\theta[\phi_{i,i+1},\tilde\delta_i\phi] = 0,
\end{equation}
which is why any terms at $B_\Upsilon$ in \eqref{Equation: tilde delta S} can be neglected. The second equality in \eqref{Equation: tilde delta S} comes from considering the action of $\mathbb{Z}_2+\mathcal{C}$, where here $\mathbb{Z}_2$ refers to reflecting everything across $\Sigma\cup\bar\Sigma$. Because $\tilde\delta\phi$ picks up a minus sign from this transformation, and additionally the orientations of the integrals are flipped, we have
\begin{equation}
    \int_{\mathcal{M}^+\cup\bar\Sigma^+}\theta[\phi_i,\tilde\delta_i\phi] = \int_{\mathcal{M}^-\cup\bar\Sigma^-}\theta[\phi_i,\tilde\delta_i\phi],
\end{equation}
from which \eqref{Equation: tilde delta S} follows.

Note that the holographic dictionary implies
\begin{equation}
    \int_{\mathcal{M}^-} \theta[\phi,\delta\phi] = \int_{\mathcal{M}^-} \dd{\tau}\dd[d-1]{x} \delta\lambda(\tau,x)\cdot \mathcal{O}(\tau,x).
\end{equation}
Since the change in boundary conditions $\delta\lambda$ for $\delta_i\phi$ and $\tilde\delta_i\phi$ at $\mathcal{M}^-$ differ by a minus sign, we can write
\begin{equation}
    -\int_{\mathcal{M}^-} \theta[\phi_i,\tilde\delta_i\phi] = \int_{\mathcal{M}^-} \theta[\phi_i,\delta_i\phi].
\end{equation}
Also, \eqref{Equation: boundary D general} implies that
\begin{equation}
    -\int_{\bar\Sigma^-} \theta[\phi_i,\tilde\delta_i\phi] = \int_{\bar\Sigma^-} \theta[\phi_i,\delta_i\phi].
\end{equation}
Note that we may write $\bar\Sigma$ instead of $\bar\Sigma^-$ for the range of integration on the right-hand side, because the integrand is single-valued there. Therefore, 
\begin{equation}
    -\frac12\tilde\delta_i S[\phi_i] = \int_{\mathcal{M}^-\cup\bar\Sigma}\theta[\phi_i,\delta_i\phi].
    \label{Equation: Uhlmann phase 2}
\end{equation}

Using \eqref{Equation: Uhlmann phase 2} in \eqref{Equation: Uhlmann phase 1}, one finds
\begin{equation}
    \gamma = -i\lim_{n\to\infty}\sum_{i=1}^n \int_{\mathcal{M}\cup\bar\Sigma}\theta[\phi_i,\delta_i\phi].
\end{equation}
The limit $n\to \infty$ may be carried out by replacing the sum by an integral. To be precise, we can write $\gamma=\oint_C a$, where
\begin{equation}
    a = -i\int_{\mathcal{M}^-\cup\bar\Sigma} \theta
\end{equation}
is a 1-form on the space of sources. This can be seen from the fact that it depends on the bulk field configuration and linearly on the variation of the bulk field configuration, so clearly it is a 1-form on the space of bulk field configurations. By pulling back this 1-form through the holographic map from boundary sources to bulk field configurations, we obtain the desired 1-form on the space of sources. $\gamma=\oint_Ca$ is the Uhlmann phase of the curve $C$.

Note that we are free to redefine $a\to a + \delta \Lambda$, where $\Lambda$ is any function on field space, and $\delta$ denotes an exterior derivative on field space. This is allowed since, by Stokes' theorem on field space, $\oint_C\delta \Lambda = 0$ for any $\Lambda$, and so the Uhlmann phase $\gamma$ is unchanged. We will use this redefinition to put $a$ in a slightly more natural form. Let
\begin{equation}
    \Lambda[\phi] = i\int_{\mathcal{N}^-} L[\phi].
\end{equation}
We have
\begin{equation}
    \delta \Lambda = i\int_{\mathcal{N}^-} \delta L = i\int_{\mathcal{N}^-}\dd{\theta} = i\int_{\mathcal{M}^-\cup\bar\Sigma\cup\Sigma} \theta.
\end{equation}
Thus, redefining $a\to a + \delta \Lambda$ yields
\begin{equation}
    a = i\int_\Sigma \theta.
    \label{Equation: Uhlmann phase connection}
\end{equation}

Before finishing this section, recall that one may add a boundary term $S_B=\int_B D$ to the action. The effect of such an addition is to change our final expression for $a$ by
\begin{equation}
    a \to a + i\delta\qty(\int_\Sigma D).
\end{equation}
But since this change is of the form $a\to a + \delta \Lambda$, it has no effect on the value of the Uhlmann phase. We previously mentioned that it might not be possible to satsify \eqref{Equation: boundary D general} by choosing the same $S_B$ for all curves of states. It should be clear now that this is of no consequence, because different choices of $S_B$ do not affect $\gamma$.

\section{Symplectic form of the entanglement wedge}
\label{Section: Symplectic form of the entanglement wedge}
In the last section, we obtained the Uhlmann phase along a curve of reduced density matrices in a subregion $A$ corresponding to boundary sources $\lambda$. We found that it was given by the integral of the 1-form $a$ around that curve, where $a$ is given in \eqref{Equation: Uhlmann phase connection}. In this section, we will treat $a$ as a connection on the space of sources, for which the Uhlmann phase is the holonomy. From this point of view, $a\to a + \delta\Lambda$ is just a gauge transformation.

Let us compute the curvature of this connection. Using $\delta$ to again denote an exterior derivative on the space of sources, the curvature is given by
\begin{equation}
    \Omega = \delta a = i\delta\qty(\int_\Sigma \theta).
\end{equation}
Since the $\delta$ is outside of the integral, we technically have to worry about field variations under which the location of the range of integration changes. In particular, the range of integration is determined dynamically by the fields, since $\Upsilon$ is the HRT surface. However, because theories of gravity are diffeomorphism invariant, we can always choose a gauge in which the range of integration is fixed. We will do this for now for simplicity.\footnote{If one wanted to consider the situation without this gauge-fixing, one would have to introduce degrees of freedom which track the location of $\Sigma$. This would lead to a formalism reminiscent of the `extended phase space' of~\cite{Donnelly:2016auv,Donnelly:2016rvo,Speranza:2017gxd}. However, an important difference is as follows. In that paper, the fields $\phi$ and the location of $\Sigma$ were more or less taken to be independent. In our setup, this is very much not the case, because $\partial\Sigma$ is dynamically determined by the fields.} One has
\begin{equation}
    \Omega = i\int_{\Sigma} \delta \theta.
\end{equation}
The components of this 2-form with respect to two particular field variations $\delta_1\phi,\delta_2\phi$ is given by
\begin{equation}
    i\int_\Sigma \omega[\phi,\delta_1\phi,\delta_2\phi],
\end{equation}
where $\omega$ is defined in \eqref{Equation: omega}. Therefore, the curvature of the Uhlmann phase is equal to the integral of $i\omega$ over $\Sigma$.

It remains to consider the Lorentzian continuation of this result. Upon Wick rotation of the fields to Lorentzian signature, we have $i\int_\Sigma\omega\to\int_\Sigma\omega$, because there are an odd number of time derivatives in $\omega$. Thus, the curvature of the Uhlmann phase is given by
\begin{equation}
    \Omega = \int_\Sigma \omega_{\text{Lor}},
\end{equation}
where $\omega_{\text{Lor}}$ denotes $\omega$ evaluated on the Lorentzian fields. According to the covariant phase space formalism, $\Omega$ is exactly the symplectic form of the domain of dependence of $\Sigma$. Since $\Sigma$ is bounded by $A$ and the HRT surface $\Upsilon$, the domain of dependence of $\Sigma$ is the entanglement wedge of $A$. Thus, we come to the main result of the paper: 
\begin{quote}
    \large The curvature of the Uhlmann phase is holographically dual to the symplectic form of the entanglement wedge.
\end{quote}

\subsection{Subregion deformations and edge modes}
\label{Section: edge modes}

Until now, we have considered the Uhlmann phase for a curve of density matrices in a fixed boundary subregion, but with varying sources. In this section, we will consider the case where we fix the sources, and vary the subregion.

Let $A(t)$, $0\le t \le 1$ be a closed smooth curve in the space of boundary subregions. We assume for simplicity that this curve takes the form $A(t) = G_t(A)$, where $G_t$ is a curve of boundary diffeomorphisms, and $A$ is a fixed subregion. For fixed sources $\lambda$, we have a density matrix for each value of $t$ given by reduction of the pure state $\ket{\lambda}$ to $A(t)$:
\begin{equation}
    \rho(t) = \tr_{\overline{A(t)}}\ket{\lambda}\bra{\lambda}.
\end{equation}
We wish to compute the Uhlmann phase along this curve, but there is an obstruction to this, in that the density matrices for different values of $t$ act on different Hilbert spaces $\mathcal{H}_{A(t)}$. In order to proceed, one must find appropriate maps from each of these Hilbert spaces to a common one, and it is not immediately obvious which maps these should be.

It has been argued by various authors~\cite{Banerjee:2011mg,Faulkner:2015csl,Faulkner:2016mzt,Lewkowycz:2018sgn,Czech:2019vih} that infinitesimal deformations of the boundary subregion may alternatively be thought of as being sourced by appropriate insertions of the stress-tensor. In particular, if we want to understand the change induced by a deformation generated by the vector field $\zeta$, one may insert
\begin{equation}
    \lie_\zeta g_{ab} \,T^{ab}
\end{equation}
in the boundary state, where $g_{ab}$ is the boundary metric, and $T^{ab}$ is the boundary stress-tensor. One may view this as a change $\lie_\zeta g_{ab}$ in the source of the operator $T^{ab}$. We will write $\lambda\to\lambda+\delta_\zeta\lambda$ to represent this change.

With this interpretation one can construct density matrices acting on the same Hilbert space. In particular one obtains infinitesimally close density matrices
\begin{align}
    \rho &= \tr_{\bar{A}} \ket{\lambda}\bra{\lambda},\\
    \rho' &= \tr_{\bar{A}} \ket{\lambda+\delta_\zeta\lambda}\bra{\lambda+\delta_\zeta\lambda},
\end{align}
which both act on $\mathcal{H}_A$. By integrating this construction along the whole curve of subregions $A(t)$, we get a curve of density matrices
\begin{equation}
    \rho(t) = \tr_{\bar{A}}\ket{\lambda(t)}\bra{\lambda(t)},
\end{equation}
all acting on $\mathcal{H}_A$. Here $\lambda(t)$ obeys $\lambda(0)=\lambda$ and
\begin{equation}
    \dv{\lambda(t)}{t} = \delta_\zeta\lambda,
\end{equation}
where $\zeta$ is the infinitesimal vector field that generates evolution along the curve of diffeomorphisms $G_t$.

With this prescription, we can now compute the Uhlmann phase, and it should be clear that this is just a special case of what we have been considering previously, because the subregion is fixed to $A$, and we are varying the sources.

Diffeomorphism invariance makes it easy to compute the bulk fields corresponding to these sources. Let $\phi$ be the bulk field configuration matching the boundary conditions set by $\lambda$, and let $H_t$ be a curve of diffeomorphisms acting on the bulk with the property that $H_t$ restricted to the boundary is equal to $G_t$. Then $\phi(t)=H_t^*\phi$ is the bulk field configuration matching the boundary conditions set by $\lambda(t)$. The field variation along this curve is given by $\delta\phi = \lie_V \phi$, where $V$ is the bulk vector field corresponding to infinitesimal evolution along $H_t$.

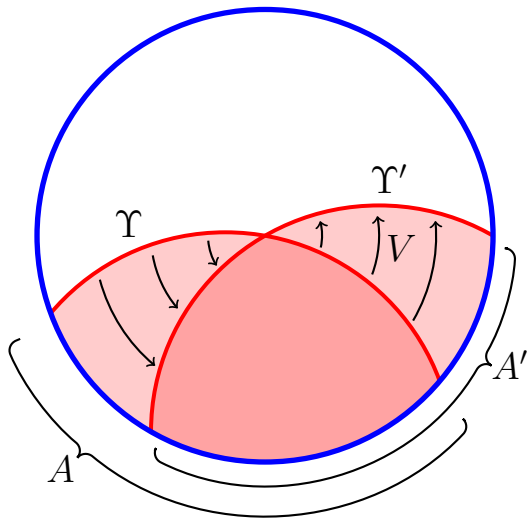
\begin{figure}
    \centering
    \begin{tikzpicture}[thick,font=\large]
        \begin{scope}
            \clip (0,0) circle (3);
            \fill[red,opacity=0.2] (-60:3) circle (3);
            \fill[red,opacity=0.2] (-100:3) circle (3);
            \draw[red,line width=1.5pt] (-60:3) circle (3);
            \draw[red,line width=1.5pt] (-100:3) circle (3);
        \end{scope}
        \draw[blue,line width=2pt] (0,0) circle (3);
        \draw[->] (-12:0.75) arc (-12:15:0.75);
        \draw[->] (-20:1.5) arc (-20:10:1.5);
        \draw[->] (-30:2.25) arc (-30:5:2.25);
        \draw[<-] (-148:0.75) arc (-148:-175:0.75);
        \draw[<-] (-140:1.5) arc (-140:-170:1.5);
        \draw[<-] (-130:2.25) arc (-130:-165:2.25);

        \draw[rounded corners] (-3:3.2) -- (-3:3.3) arc (-3:-30:3.3) -- (-30:3.4);
        \draw[rounded corners] (-117:3.2) -- (-117:3.3) arc (-117:-30:3.3) -- (-30:3.4);
        \node at (-28:3.65) {$A'$};
        \draw[rounded corners] (-43:3.6) -- (-43:3.7) arc (-43:-130:3.7) -- (-130:3.8);
        \draw[rounded corners] (-157:3.5) -- (-157:3.7) arc (-157:-130:3.7) -- (-130:3.8);
        \node at (-131:4.1) {$A$};

        \node at (25:1.8) {$\Upsilon'$};
        \node at (175:1.8) {$\Upsilon$};

        \node at (-5:1.8) {$V$};
    \end{tikzpicture}
    \caption{A deformation of the boundary subregion $A\to A'$ leads to a corresponding deformation of the HRT surface $\Upsilon\to \Upsilon'$, generated by a bulk vector field $V$.}
    \label{Figure: HRT deformation}
\end{figure}

Each boundary subregion $A(t)$ has an associated HRT surface $\Upsilon(t)$ in the bulk. Recall that we have fixed the gauge such that $\Sigma$ must have its boundary at the HRT surface. In order for $H_t$ to be consistent with this gauge choice, it must be the case that $H_t(\Upsilon) = \Upsilon(t)$. Thus, the vector field $V$ at $\Upsilon$ generates the deformation of the HRT surface $\Upsilon(t)$. This is depicted in Figure~\ref{Figure: HRT deformation}.

Let us compute the components of the symplectic form with respect to this deformation, i.e.
\begin{equation}
    \Omega[\phi,\lie_V\phi,\delta\phi] = \int_\Sigma\omega[\phi,\lie_V\phi,\delta\phi].
\end{equation}
There is a well-known formula that expresses this quantity in terms of an integral over $\partial\Sigma$ that has appeared, for example, in~\cite{Wald:1993nt}. We will briefly run over the argument here. The Lie derivative of the Lagrangian density with respect to any bulk vector field $\xi$ is
\begin{equation}
    \lie_\xi L[\phi] = \dd(\iota_\xi L[\phi]).
\end{equation}
Under the assumption that $L$ is covariantly constructed from the fields $L$, the Lie derivative of $L$ may also be written as the variation of $L$ with respect to a variation of the bulk fields by Lie derivatives, $\delta\phi=\lie_\xi\phi$. Thus, using $\delta L = \dd{\theta}$, we have
\begin{equation}
    \lie_\xi L[\phi] = \dd{\theta[\phi,\lie_\xi\phi]}.
\end{equation}
Equating the two right-hand sides above yields the result that
\begin{equation}
    j_\xi[\phi] = \iota_\xi L[\phi] - \theta[\phi,\lie_\xi\phi]
\end{equation}
is a closed form. It is the (Hodge dual of) the Noether current associated with $\xi$. Since it is closed for any $\xi$, the results of~\cite{Wald:1990} imply that it is in fact exact, and we write $j_\xi=\dd{q_\xi}$. The form $q_\xi$ is the (Hodge dual of) the Noether charge density associated with $\xi$. 

The commutator of the variations $\lie_V\phi$ and $\delta\phi$ acts as
\begin{equation}
    \begin{aligned}
        \phi &\to \phi + [\lie_V,\delta]\phi \\
             &= \phi + \lie_V(\delta\phi) - \delta(\lie_V\phi) \\
             &= \phi - \lie_{\delta V}\phi.
    \end{aligned}
\end{equation}
Therefore, we have $[\lie_V,\delta]=-\lie_{\delta V}$. 

Using the above we may write
\begin{equation}
    \begin{aligned}
        \Omega[\phi,\lie_V\phi,\delta\phi] &= \int_\Sigma \lie_V(\theta[\phi,\delta\phi]) - \delta(\theta[\phi,\lie_V\phi]) + \theta[\phi,\lie_{\delta V}\phi]\\
                                           &= \int_\Sigma \dd(\iota_V\theta[\phi,\delta\phi]) + \iota_V\dd{\theta[\phi,\delta\phi]} - \delta(\iota_V L[\phi] - \dd{q_V[\phi]}) + \iota_{\delta V} L[\phi] - \dd{q_{\delta V}[\phi]} \\
                                           &= \int_\Sigma \dd(\delta q_V[\phi] - q_{\delta V}[\phi] + \iota_V\theta[\phi,\delta\phi]) + \iota_V\delta L[\phi] - \delta(\iota_V L[\phi]) +\iota_{\delta V} L[\phi] \\
                                           &= \int_{\partial\Sigma} \delta q_V[\phi] - q_{\delta V}[\phi] + \iota_V\theta[\phi,\delta\phi].
    \end{aligned}
\end{equation}
Thus, the components of the curvature of the Uhlmann phase with respect to a subregion deformation reduce to a boundary integral at $\partial\Sigma$.

This feature of the covariant phase space formalism is not unique to our situation, and has been observed many times before~\cite{Regge:1974,Brown:1986,Wald:1990,Wald:1993nt,Wald:1999wa,Barnich:2000zw,Barnich:2001jy,Compere:2018aar}. However, there has previously not been much reason to restrict the form which $V$ can take.\footnote{Sometimes, boundary conditions have been imposed at $\partial\Sigma$. Then $V$ must preserve these boundary conditions. Usually, however, the boundary conditions are somewhat arbitrary, and no a priori justification for them is given. } In our case, $V$ is much more constrained -- it must be a vector field representing an infinitesimal deformation of one HRT surface to another nearby HRT surface.\footnote{In the language of the previous footnote, we now have the non-arbitrary boundary condition that $\Upsilon$ must be an HRT surface.} Thus, we get a classical degree of freedom, living at $\partial\Sigma$, for each such deformation. Such degrees of freedom are referred to as edge modes.

For some deformations, it may be the case that we can write
\begin{equation}
    \Omega[\phi,\lie_V\phi,\delta\phi] = \delta H_V,
\end{equation}
where $H_V=H_V[\phi]$ is some function on field space. Then we say that the deformation is integrable, and may view $H_V$ as the Hamiltonian generating the deformation. It is an interesting question to ask which deformations are integrable. This will be the topic of future work, but for now let us simply comment that this question is intimately associated with the conformal symmetry of the boundary theory. Indeed, consider the case where the subregion $A$ contains the entire boundary. Then the HRT surface is empty, and $\Omega[\phi,\lie_V\phi,\delta\phi]$ is expressible entirely in terms of the fields on the boundary, and in this case the question reduces to asking when the curve of states $\ket{G_t^*\lambda}$ can be written as $e^{iH t}\ket{\lambda}$, for some boundary operator $H$. Of course, the answer is that this is the case when $G_t$ is a conformal transformation. Then $H$ is simply the generator of that conformal transformation. When $A$ is a proper subregion of the boundary, the situation becomes more complicated due to the fact that $\Omega[\phi,\lie_V,\delta\phi]$ contains terms at the HRT surface. But clearly the conformal transformations are a good place to start.

Because $\Omega[\phi,\lie_V\phi,\delta\phi]$ contains contributions at the HRT surface, it is in principle possible to use the Uhlmann phase arising from deformations of the boundary subregion $A$ to measure the fields near the HRT surface, including the Riemann curvature. Similar conclusions were drawn in~\cite{Czech:2019vih}. However, in that paper the authors discussed a different type of phase, which they called the modular Berry phase. It would be interesting to understand the relationship between the Uhlmann phase and the modular Berry phase in the holographic context.

\subsection{Resolution of the boundary ambiguity}
\label{Section: ambiguity resolution}

Recall from the Introduction that there is an ambiguity in the definition of $\theta$. In particular one is allowed to modify $\theta$ by any exact form. This changes the integral $\int_\Sigma \theta$ by a boundary term at $\partial\Sigma$. It is clear that one must fix this ambiguity if one is to understand the degrees of freedom near the boundary.

In Section~\ref{Section: holographic Uhlmann phase}, we used this freedom to enforce the condition
\begin{equation}
    \int_{B_\Upsilon} \theta[\phi_{i,i+1},\delta\phi] = 0.
    \label{Equation: theta condition 2}
\end{equation}
We will now show that this condition fixes the boundary ambiguity at $\Upsilon$.

Suppose \eqref{Equation: theta condition 2} holds, and we attempt to modify $\theta\to\theta + \dd{K}$. Then we would have
\begin{equation}
    0 = \int_{B_\Upsilon} \theta[\phi_{i,i+1},\delta\phi] \to \int_{B_\Upsilon} \theta[\phi_{i,i+1},\delta\phi] + \int_{\partial B_\Upsilon} K[\phi_{i,i+1},\delta\phi].
\end{equation}
Clearly \eqref{Equation: theta condition 2} only continues to hold if the integral over $\partial B_\Upsilon$ vanishes. $\partial B_\Upsilon$ has two components $\Upsilon^\pm = \partial\bar\Sigma^\pm\cap B_\Upsilon$. We thus require
\begin{equation}
        0 = \int_{\Upsilon^+} K[\phi_{i,i+1},\delta\phi] - \int_{\Upsilon^-} K[\phi_{i,i+1},\delta\phi].
\end{equation}
In order for our expression of the Uhlmann curvature to be valid in all situations, we want this to hold for \emph{any} possible $\phi_{i,i+1}$ and $\delta\phi$ which obey the equations of motion everywhere but at $\bar\Sigma$. The contributions of these fields at $\Upsilon^-$ and $\Upsilon^+$ are essentially independent of one another, and freely specifiable in the above. Thus, we need
\begin{equation}
    0 = \int_{\Upsilon^\pm} K[\phi,\delta\phi]
    \
\end{equation}
for any $\phi,\delta\phi$. In the limit as $B_\Upsilon$ tightly encloses $\Upsilon$, we can replace $\Upsilon^\pm\to\Upsilon$, and so obtain
\begin{equation}
    0 = \int_\Upsilon K[\phi,\delta\phi].
\end{equation}
But note that $\theta\to\theta + \dd{K}$ for such a $K$ implies
\begin{equation}
    \int_\Sigma \theta \to \int_\Sigma \theta + \int_\Upsilon K[\phi,\delta\phi] = \int_\Sigma \theta.
\end{equation}
Here we are assuming that the ambiguity is fixed at asymptotic infinity in some other way, so there is no contribution from $K$ there. Thus, any change $\theta\to\theta + \dd{K}$ that respects \eqref{Equation: theta condition 2} must lead to no change in $\int_\Sigma\theta$. So, this condition does indeed fix the ambiguity at the HRT surface.

The condition \eqref{Equation: theta condition 2} is formulated as a Euclidean expression. However, for practical purposes it would be more convenient to be able to state it in terms of the fields in the Wick rotated Lorentzian bulk spacetime. Also, it would be useful to see if \eqref{Equation: theta condition 2} could be understood in a simple and convenient way for some basic examples, such as pure Einstein gravity. We leave these and other questions for future work.

\section{Conclusion}
\label{Section: Conclusion}

In this paper we have argued that the holographic dual of the symplectic form in an entanglement wedge is the curvature of the Uhlmann phase for states reduced to the corresponding boundary subregion. Let us briefly speculate on some consequences and possible future applications of this result.

First, our result gives a specific operational context to the concept of classically emergent physics in a subregion that was previously somewhat absent: classical bulk subregion physics emerges in measurements of the Uhlmann phase. It is important to point out that the Uhlmann phase is a genuine observable, as has been argued in principle~\cite{Operational}, and has recently has been confirmed in practice~\cite{Observation}. It would be useful to figure out more of the details of this context.

Second, we would like to more fully understand the resolution of the boundary ambiguity for the symplectic form given in this paper, and its implications for edge modes. For example, the edge modes have been used to try to understand black hole entropy~\cite{Haco:2018ske,Haco:2019ggi}, and it would be worthwhile to see if the methods used in those papers are consistent with our results.

Third, starting from the quantum mechanical description of a complete holographic system, our construction resulted in a classical phase space for the degrees of freedom in the entanglement wedge. It is natural to attempt to run this backwards, i.e.\ to quantise this phase space. One would then obtain an `effective' Hilbert space for the entanglement wedge. In the original system, all the states in the entanglement wedge had to be mixed, because of entanglement in the CFT. However, the effective entanglement wedge Hilbert space is clearly made up of \emph{pure} states. Hence, by studying such a quantisation, one should be able to learn about what it means to have a pure state in a gravitational subregion. This would be of particular interest in the case of a black hole spacetime, where the entanglement wedge is chosen to coincide with the black hole exterior. The pure states in the effective Hilbert space might then reasonably be called black hole microstates.

Fourth, the calculation presented in this paper applies at leading order for large $N$. It would be interesting to try to understand the subleading corrections, where the condition on the HRT surface is supposed to be changed from extremising the area, to extremising the generalised entropy~\cite{Faulkner:2013ana}.

Fifth, in~\cite{Belin:2018bpg} the holographic Berry curvature was used to investigate the complexity of holographic states. It may be possible to use our results to extend that analysis to holographic subregion complexity, which has previously been explored in~\cite{Alishahiha:2015rta,Ben-Ami:2016qex,Carmi:2017jqz,Carmi:2017ezk,Chen:2018mcc,Agon:2018zso,Abt:2018ywl}.

Finally, Uhlmann phases have been used to classify phases of condensed matter systems~\cite{Fermions}. It would be interesting to see if our expression for the Uhlmann phase could be used in a similar way, in the cases where the systems have holographic duals.

\section*{Acknowledgements}
\addcontentsline{toc}{section}{\protect\numberline{}Acknowledgements}

I wish to thank Joan Camps, Kelley Kirklin, Jo\~ao Melo, Malcolm Perry, G\'abor S\'arosi, David Skinner, and Aron Wall for insightful discussions and helpful comments. I am also grateful for the stimulating environment provided by the QIQG5 workshop at Davis. This work was supported by a grant from STFC.

\printbibliography

\end{document}